\documentclass[10pt,preprint,showpacs,preprintnumbers,amsmath,amssymb,floatfix]{revtex4}
\usepackage{graphicx,epsfig}
\usepackage{subfigure}
\newcommand\e{\mathrm{e}}
\newcommand\ele{\mathrm{l}}
\newcommand\ere{\mathrm{r}}

\newcommand\im{\mathrm{i}}
\newcommand\imag{\mathrm{Im}}

\newcommand\num{\mathrm{num}}
\newcommand\phai{\mathrm{pi}}
\newcommand\sr{\mathrm{sr}}
\newcommand\real{\mathrm{Re}}
\newcommand\ret{\mathrm{ret}}
\newcommand\s{\mathrm{s}}
\newcommand\sca{\mathrm{S}}
\newcommand\ten{\mathrm{T}}

\newcommand\ua{\mathrm{ua}}

\newcommand\D{\mathrm{d}}

\newcommand\Mpc{\mathrm{Mpc}}
\newcommand\Pl{\mathrm{Pl}}

\begin{document}

\title{Computation of inflationary cosmological perturbations in chaotic inflationary scenarios using the phase-integral method}

\author{Clara Rojas$^{1,}$}
\email{clararoj@gmail.com}
\author{V\'ictor M. Villalba$^{2,}$}
\email{villalba@ivic.ve}
\affiliation{$^1$Departamento de F\'isica Aplicada IVIC/M\'erida, M\'erida 5107, Venezuela}
\affiliation{$^2$Centro de F\'isica IVIC Apdo 21827, Caracas 1020A, Venezuela}

\date{\today}

\begin{abstract}
The phase-integral approximation devised by Fr\"oman and Fr\"oman, is used for computing cosmological  perturbations in the quadratic chaotic inflationary model. The phase-integral formulas for the scalar and  tensor power spectra are explicitly obtained up to fifth order of the phase-integral approximation. We show that, the phase integral  gives a very good approximation for the shape of the power spectra associated with scalar and tensor perturbations as well as the spectral indices. We find that the accuracy of the phase-integral approximation compares favorably with the numerical results and those obtained using the slow-roll  and uniform approximation methods.
\end{abstract}


\maketitle

\section{Introduction}

The study of the spectrum of  anisotropies of the  cosmic microwave background radiation and inhomogeneities in the large scale structure of the universe provides key elements in the study of the early universe. The results reported by WMAP favor inflation \cite{komatsu:2009,spergel:2007,spergel:2003} over other cosmological scenarios. The present and future observational data will permit us to validate and discriminate among different inflationary models.  According to WMAP5 data the power-law inflation, the hybrid inflation and quartic chaotic inflationary models $\lambda \phi^4$ are ruled out, while the quadratic chaotic inflationary model $m^2\phi^2$ agreed with the observational data \cite{komatsu:2009,kinney:2008}.

In order to compare with observations,  we should be able to obtain very accurate results for the predicted power spectrum of primordial perturbations for a variety of inflationary scenarios. In general, most of the inflationary models are not exactly solvable and approximate or numerical methods are mandatory in the computation of the scalar and power spectra.  Traditionally, the method of approximation applied in inflationary cosmology is the slow-roll approximation \cite{stewart:1993}, which produces reliable results in inflationary models with smooth potentials,  but cannot
be improved  on a simple way beyond the leading order. Recently, some authors have applied alternative approximations, such as the WKB method with the Langer modification \cite{langer:1937,martin:2003A,casadio:2005A}, the Green function method \cite{stewart:2001}, and the improved WKB method \cite{Casadio2005C,Casadio2005B}.

Habib \textit{et al} \cite{habib:2002, habib:2004,habib:2005B}  have successfully applied the uniform approximation method in the calculation of the scalar and tensor power spectra and the corresponding spectral indices for the quadratic and quartic chaotic inflationary models, showing that the the uniform approximation gives more accurate results than the slow-roll approximation. Casadio \textit{et al} \cite{Casadio2006} have applied  the method of comparison equation to study cosmological perturbations during inflation.  The comparison method is based on  the uniform approximation proposed by Dingle \cite{Dingle} and Miller \cite{Miller} and thoroughly discussed by Berry and Mount \cite{Berry}.

Recently \cite{rojas:2007b,rojas:2007c}, the phase-integral
approximation \cite{froman:1965,froman:1996,froman:2002} has been
applied to the calculation of the power spectra and spectral indices
in the power-law inflationary model, showing that the phase-integral
method gives results which are comparable or better than those
obtained using slow-roll or the uniform approximation. It is purpose
of this paper to  compute approximate solutions for the scalar and
tensor power spectra and their corresponding spectral indices for
some chaotic inflationary models  with the help of the
phase-integral approximation method. We show that for the quadratic
chaotic inflationary model the fifth-order phase integral
approximation gives more accurate results than those obtained using
the WKB or uniform approximation methods.  For the quartic chaotic
inflationary model we also obtain very accurate results but we do
not report them since this model is ruled out by the observational
data \cite{komatsu:2009}.

The article is structured as follows: In Sec. II we apply the
phase-integral approximation to the  chaotic quadratic inflationary
model,  we numerically solve the equation governing the scalar and
tensor perturbations and compare the results for the power-spectra
obtained using the phase-integral approach with those computed with
the slow-roll and uniform-approximation methods. In Sec. III we
summarize our results.

\section{Phase-integral approximation for the power spectrum in the  $\phi^2$ chaotic inflation}

In this section we discuss the application of the phase-integral
approximation to the computation of the power spectrum in the
chaotic inflationary $\frac{1}{2}m^2\phi^2$ model. We apply the
phase-integral approximation in the study of the evolution of the
mode $k$ equations for the scalar and tensor perturbations in order
to compute the scalar and tensor power spectra.  Using the
fifth-order phase integral approximation we compute the scalar and
tensor power spectra and their corresponding spectral indices. A
detailed description of the method is given in reference
\cite{rojas:2007b,rojas:2007c}.

\subsection{The model}
The chaotic inflationary model was introduced by Linde
\cite{linde:1983A,linde:1983B}, he proposed that the preinflationary
universe was chaotic  which means that the fields would take
different values in different points of the space following a random
pattern and inflation will occur in virtually any universe that
begins in a chaotic, high energy state and has a scalar field with
unbounded potential energy. The simplest form  of the inflaton
potential $V(\phi)$ in a chaotic model is given by the quadratic
potential

\vspace{-0.4cm}
\begin{equation}
\label{ch2_V}
V(\phi)=\frac{1}{2} m^2 \phi^2,
\end{equation}
giving as a result a free scalar field with mass $m$.

\subsection{Equations of motion}\label{ch2_moveq}
In an inflationary universe driven by a scalar field, the equations
of motion for the inflaton $\phi$ and the Hubble parameter $H$ are
given by

\begin{eqnarray}
\label{ch2_ddotphi}
\ddot{\phi}&+&3H\dot{\phi}=-\frac{\partial V(\phi)}{\partial\phi},\\
\label{ch2_H^2}
H^2&=&\frac{1}{3M_\Pl^2}\left[V(\phi)+\frac{1}{2}\dot{\phi}^2\right],
\end{eqnarray}
where the dots indicate derivatives with respect to physical time
$t$. In the quadratic chaotic inflationary model the Eqs.
\eqref{ch2_ddotphi} and \eqref{ch2_H^2} are not exactly solvable in
closed form; they can be solved numerically or using the slow-roll
approximation. In the slow-roll approximation \cite{liddle:2000} we
consider that the scalar field $V(\phi)$ varies very slowly
$\frac{1}{2}\dot{\phi}^{2}\ll V(\phi)$. Using this approximation, we
obtain that  Eq. \eqref{ch2_ddotphi} and Eq.\eqref{ch2_H^2} for the
quadratic chaotic inflationary model reduce to \cite{copeland:2004}

\begin{eqnarray}
\label{phi_phi2}
3H\dot{\phi}&\simeq& -m^2\phi,\\
\label{H_phi2}
H&\simeq&\frac{m\phi}{\sqrt{6} \,M_\Pl}.
\end{eqnarray}
Using Eq. \eqref{phi_phi2} and Eq. \eqref{H_phi2} we obtain that, in
the slow-roll approximation, the expansion factor $a(t)$ and the
inflaton field $\phi(t)$  are

\begin{eqnarray}
\label{ch2_phisr}
\phi_\sr&\simeq&\phi_i-\sqrt{\frac{2}{3}}m \,M_\Pl t,\\
\label{ch2_asr}
a_\sr&\simeq& a_i\exp\left({\frac{m\phi_i}{\sqrt{6} \,M_\Pl} t -\frac{m^2}{6} t^2}\right),
\end{eqnarray}
where $\phi_i$ is a constant of integration corresponding to the
initial value of the inflaton. Eq. \eqref{ch2_asr} shows that the
Universe expands exponentially during inflation  The slow-roll
parameter is \cite{stewart:2001}

\begin{equation}
\epsilon_1=\frac{\dot{H}}{H^2}=\frac{2\,M_\Pl^2}{\phi^2}.
\end{equation}
The inflationary epoch finishes when $\epsilon_1=1$, that is
$\phi_\textnormal{f}=\sqrt{2}\,M_\Pl$, when the scalar field starts
to oscillate. The mass $m$ of the inflation can be fixed using the
amplitude of the density fluctuations detected by  WMAP. In order to
fit $m$ to the observational data, we demand that $ m\simeq 10^{-6}$
\cite{spergel:2003}.

The equations of motion \eqref{ch2_ddotphi} and  \eqref{ch2_H^2} are
numerically integrated in the physical time $t$. We solve the system
of coupled differential equations \eqref{ch2_ddotphi} and
\eqref{ch2_H^2} with the help of the sixth-order Runge-Kutta method
\cite{gerald:1984}, which can be written as

\begin{eqnarray}
\label{ch2_system}
\ddot{\phi}&+&3\displaystyle{\frac{\dot{a}}{a}}\dot{\phi}+\frac{\partial V(\phi)}{\partial \phi}=0,\\
\nonumber
\displaystyle{\frac{\dot{a}}{a}}&-&\sqrt{6} \,M_\Pl\left[2 V(\phi)+\dot{\phi}^2\right]^{1/2}=0.
\end{eqnarray}

We choose the initial value of the inflaton as
$\phi_i=15.4\,\,M_\Pl$. Since the evolution of the inflation $\phi$
is governed by a second  a second-order differential equation, we
need to fix the initial value for the velocity of the scalar field
$\dot{\phi}_i$, which can be obtained using the slow-roll
approximation \eqref{ch2_phisr}. The initial value for  $a_i$ is
chosen as $a_i=1$, the mass of the inflaton  $m^2=1.89\times
10^{-12} \,M_\Pl^2$.  The initial condition has been selected in
order to guarantee enough inflation. In order to find the number of
e-folds, we rewrite the system of differential equations
\eqref{ch2_system} as

\begin{eqnarray}
\label{ch2_sys:N,phi}
\ddot{\phi}&+&3\dot{N}\dot{\phi}+\frac{\partial V(\phi)}{\partial \phi}=0,\\
\nonumber
\dot{N}&=&\sqrt{6} \,M_\Pl\left[2 V(\phi)+\dot{\phi}^2\right]^{1/2}.
\end{eqnarray}

\begin{figure}[htbp]
\begin{center}
\includegraphics[scale=0.465]{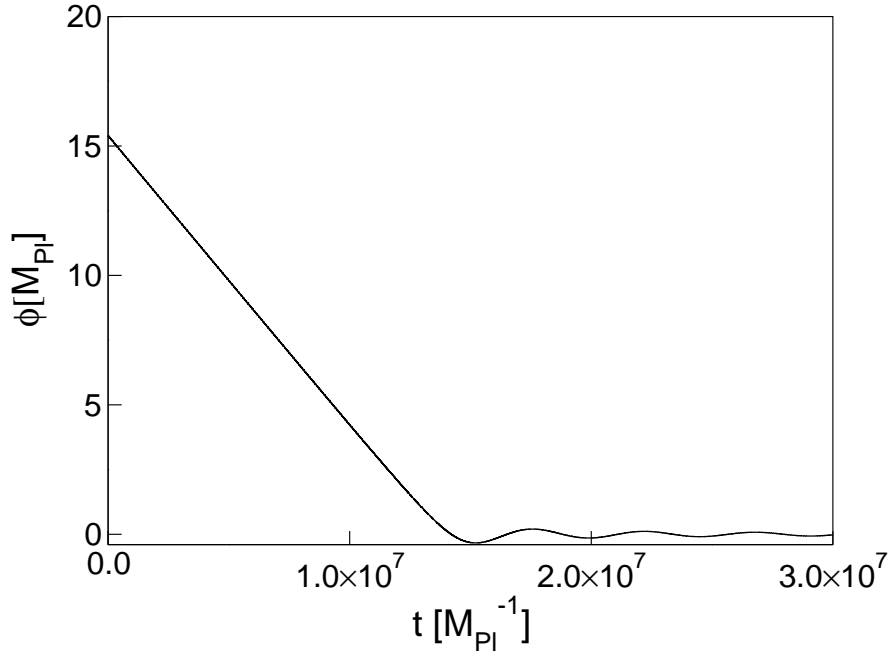}\\
\caption{\small{Evolution of the potential  $\phi$ for the chaotic inflationary model $\frac{1}{2}m^2\phi^2$ }}
\label{ch2:phicomplete}
\end{center}
\end{figure}

We find that the inflation finishes at $t_f=1.30\times 10^7
\,M_\Pl^{-1}=3.51\times 10^{-36}\s$, a result that corresponds to
$59.84$ e-folds before the scalar field starts to oscillate. Fig.
\ref{ch2:phicomplete} shows the evolution of the scalar field
$\phi$.

\vspace{1cm}

\begin{figure}
\begin{center}
\subfigure[]{
\label{ch2:a}
\includegraphics[scale=0.465]{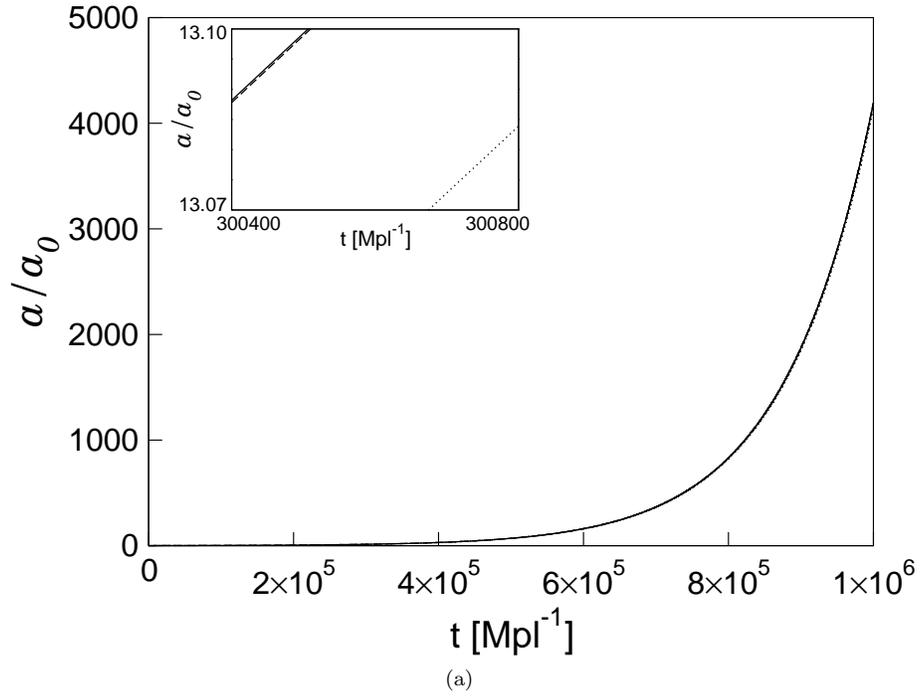}}\\
\vspace{1.2cm}
\subfigure[]{
\label{ch2:phi}
\includegraphics[scale=0.465]{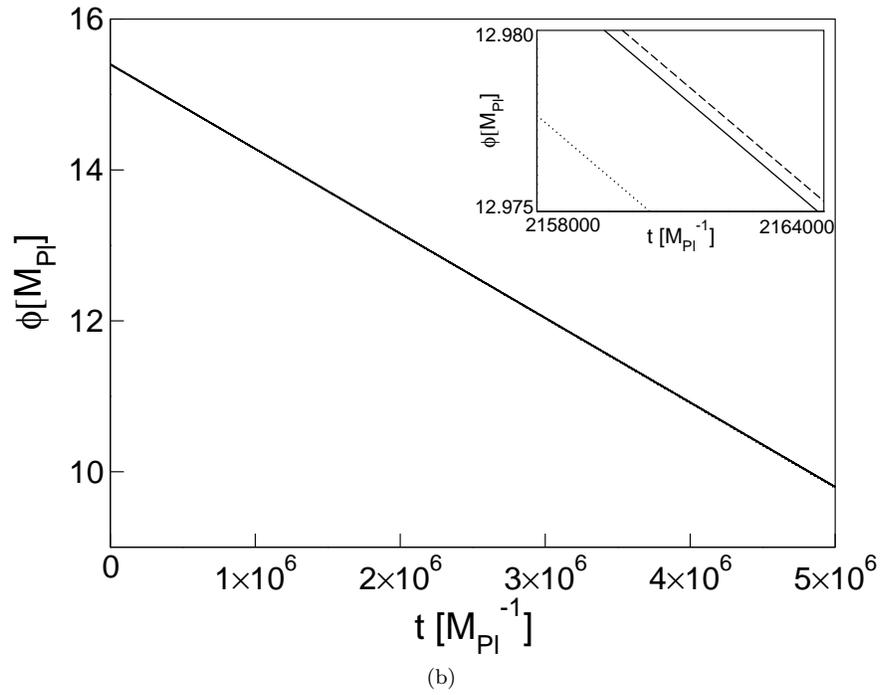}}
\caption{\small{(a) Evolution of the scale factor $a$  and (b) evolution of the inflaton  $\phi$ for the chaotic inflationary model $\frac{1}{2}m^2\phi^2$. Solid line: numerical solution; dashed line: fitting; dotted line: slow-roll approximation. The inset is an enlargement of the figure.}}
\end{center}
\end{figure}

\begin{figure}[htbp]
\begin{center}
\subfigure[]{
\label{ch2:a_1}
\includegraphics[scale=0.465]{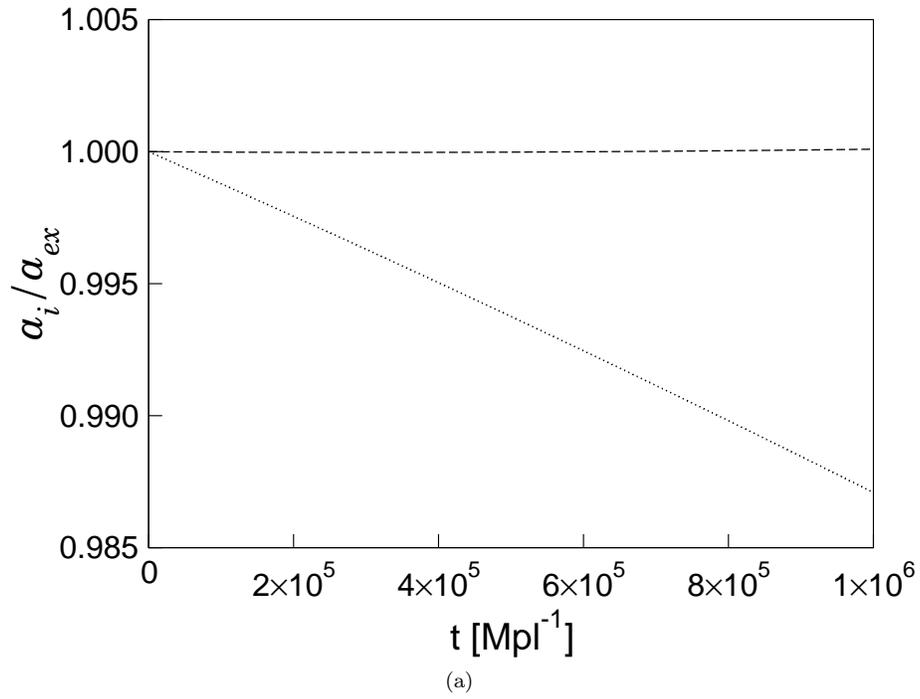}}\\
\vspace{1.2cm}
\subfigure[]{
\label{ch2:phi_1}
\includegraphics[scale=0.465]{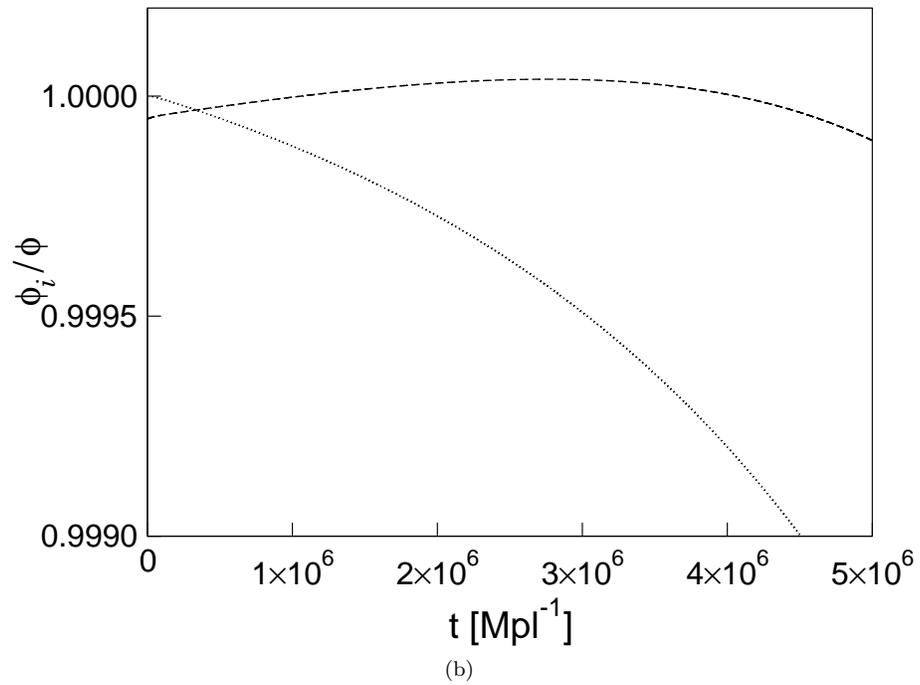}}
\caption{\small{(a) Ratio of the fit and slow-roll to the exact solution for $a(t)$: $a_i/a$  and (b) ratio of the fit and slow-roll to the exact solution for $\phi(t)$: $\phi_i/\phi$, in the chaotic inflationary model $\frac{1}{2}m^2\phi^2$. Dashed line: fitting; dotted line: slow-roll approximation.}}
\end{center}
\end{figure}

In order to apply the phase-integral approximation to higher orders
it is necessary to calculate the integrals  $\omega(z)$  in the
complex plane, therefore,  it is of help to have an analytic
expression for $a(t)$ and $\phi(t)$ which can be obtained after
fitting the numerical data. We find  that   $a(t)$ and  $\phi(t)$,
take the form

\begin{eqnarray}
\label{ch2_a_fit}
a(t)&=&a_i \exp\left(\alpha t-\beta t^2\right),\\
\label{ch2_phi_fit}
\phi(t)&=&\gamma-\sigma t,
\end{eqnarray}
where  $\alpha=8.6551\times10^{-6}\, M_\Pl$,
$\beta=3.1380\times10^{-13}\,M_\Pl^2$, $\gamma=15.3992\, M_\Pl$ and
$\sigma=1.1201\times 10^{-6} \,M_\Pl^2$.  With the help of the
expressions  \eqref{ch2_a_fit} and  \eqref{ch2_phi_fit} we obtain
that   $z_\sca$ is given by:

\begin{equation}
\label{ch2_zs}
z_\sca(t)=-\frac{a_i\gamma}{(\alpha-2\beta t)} \exp\left(\alpha t -\beta t^2\right).
\end{equation}
Fig. \ref{ch2:a} and  Fig. \ref{ch2:phi} compare the fitting with the numerical result and the slow-roll approximations for $a(t)$ and $\phi(t)$. The fitting is valid up to $t=5.00\times 10^6 \,M_\Pl^{-1}$. The inset is an enlargement of the figure. Fig. \ref{ch2:a_1} and  Fig. \ref{ch2:phi_1} show the ratio of the fit and slow-roll to exact solution for $a(t)$ and $\phi(t)$ and observe that the fitting better approximates the numerical result than the slow-roll approximation, therefore the expressions for the scalar and tensor perturbations will be constructed using the fitting $a(t)$ and $z_\sca(t)$ given by expressions  \eqref{ch2_a_fit} and \eqref{ch2_zs}, respectively.
If we use $a_\sr$ and $\phi_\sr$ in order to calculate the power spectrum, the expression we obtain  does not approach the exact result.

\subsection{Equation for the perturbations}\label{ch2_pereq}
Since the expansion factor $a$ and the field $\phi$ exhibit a simpler form in the physical time $t$ than in the conformal time $\eta$, we proceed to write the equations for the scalar and tensor perturbations in the variable $t$.  The relation between $t$ and $\eta$ is given via the equation $dt=a\,d\eta$. In this case, the equation for the perturbations can be written as

\begin{eqnarray}
\label{dotu}
\ddot{u_k}+\frac{\dot{a}}{a}\dot{u_k}+\frac{1}{a^2}\left[k^2-\frac{\left(\dot{a}\dot{z_\sca}+a\ddot{z_\sca}\right)a}{z_\sca} \right]u_k&=&0,\\
\label{dotv}
\ddot{v_k}+\frac{\dot{a}}{a}\dot{v_k}+\frac{1}{a^2}\left[k^2-\left(\dot{a}^2+a\ddot{a}\right) \right]v_k&=&0.
\end{eqnarray}

In order to apply the phase-integral approximation, we eliminate the terms $\dot{u}_k$ and $\dot{v}_k$ in Eq. \eqref{dotu} and Eq. \eqref{dotv}. We make the change of variables $u_k(t)=\frac{U_k(t)}{\sqrt{a}}$ and $v_k(t)=\frac{V_k(t)}{\sqrt{a}}$, obtaining that $U_{k}$ and $V_{k}$ satisfy the differential equations:

\begin{eqnarray}
\label{ch2_ddotUk}
\ddot{U}_k+R_\sca(k,t)U_k&=&0,\\
\label{ch2_ddotVk}
\ddot{V}_k+R_\ten(k,t)V_k&=&0,
\end{eqnarray}
with

\begin{eqnarray}
\label{ch2_RS0}
R_\sca(k,t)&=&\frac{1}{a^2}\left[k^2-\frac{\left(\dot{a}\dot{z_\sca}+a\ddot{z_\sca}\right)a}{z_\sca} \right]+\frac{1}{4a^2}\left(a^2-2a\ddot{a}\right),\\
\label{ch2_RT0}
R_\ten(k,t)&=&\frac{1}{a^2}\left[k^2-\left(\dot{a}^2+a\ddot{a}\right) \right]+\frac{1}{4a^2}\left(a^2-2a\ddot{a}\right),
\end{eqnarray}
where $U(k)$ satisfies the asymptotic conditions

\begin{eqnarray}
\label{ch2_cero_Uk}
U_k&\rightarrow&A_k \sqrt{a(t)} z_\sca(t),\quad  k\,t\rightarrow \infty\\
\label{ch2_borde_Uk}
U_k&\rightarrow&\sqrt{\frac{a(t)}{2k}}\exp{\left[-ik\eta(t)\right]}, \quad k\,t\rightarrow 0,
\end{eqnarray}
the asymptotic conditions (\ref{ch2_cero_Uk}) and (\ref{ch2_borde_Uk}) also hold for $V_k$.

\begin{figure}[htbp]
\begin{center}
\includegraphics[scale=0.465]{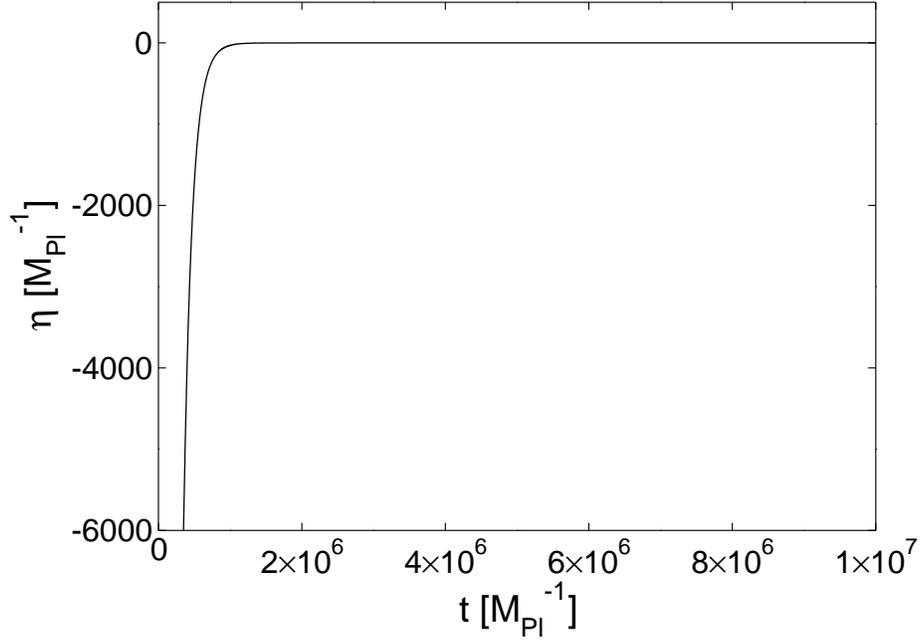}
\caption{\small{Behavior of $\eta$ as a function of the physical time $t$ for the chaotic inflationary model $\frac{1}{2}m^2 \phi^2$.}}
\label{ch2:eta}
\end{center}
\end{figure}

We now proceed to write the explicit equations for quadratic chaotic
inflation. From  Eq. \eqref{ch2_RS0} and Eq. \eqref{ch2_RT0},  with
Eq. \eqref{ch2_a_fit} and  Eq. \eqref{ch2_phi_fit} we obtain

\begin{eqnarray}
\label{ch2_RS}
R_\sca(k,t)&=&\frac{k^2}{a_i^2} \exp\left[-2t(\alpha-\beta t)\right]-\frac{\left[32\beta^2+9\left(\alpha-2\beta t\right)^4\right]}{4\left(\alpha-2\beta t\right)^2}-3\beta,\\
\label{ch2_RT}
R_\ten(k,t)&=&\frac{k^2}{a_i^2} \exp\left[-2t(\alpha-\beta t)\right]-\frac{9}{4}\left(\alpha-2\beta t\right)^2+3\beta.
\end{eqnarray}
In order to apply  the asymptotic condition \eqref{ch2_borde_Uk}, we  use of the relation between $\eta$ and $t$, which is given by:

\begin{equation}
\eta=\frac{\sqrt{\pi}}{2a_i\sqrt{\beta}}\exp\left(\frac{-\alpha^2}{4\beta}\right)\left[\mathrm{Erfi}\left(\frac{-\alpha+2\beta t}{2\sqrt{\beta}}\right)+\mathrm{Erfi}\left(\frac{\alpha-2\beta t_0}{2\sqrt{\beta}}\right)\right],
\end{equation}
where $\mathrm{Erfi}(z)$ is the imaginary error function \cite{Abramowitz}. Since the conformal time $\eta$ is defined up to an integration constant, the lower limit $t_i$ of the integral

\begin{equation}
\D\eta=\int_{t_i}^t \frac{\D t}{a(t)},
\end{equation}
is chosen in order to make  $\eta=0$ at the end of the inflationary
epoch, i.e,  $t_i= 10^7 \,M_\Pl^{-1}$. The dependence  of $\eta$ on
$t$ is shown in Fig. \ref{ch2:eta}. We can observe that as

\begin{eqnarray}
-k\,\eta\rightarrow 0         &\Rightarrow&  k\,t \rightarrow \infty,\\
-k\,\eta\rightarrow \infty  &\Rightarrow& k\,t \rightarrow 0.
\end{eqnarray}
Eq. \eqref{ch2_ddotUk} and  Eq. \eqref{ch2_ddotVk}, where
$R_\sca(k,t)$ and  $R_\ten(k,t)$ are given by Eq. \eqref{ch2_RS} and
Eq. \eqref{ch2_RT}, do not  possess exact analytic solution. In
order to solve the differential equations governing the scalar and
tensor perturbations in the physical time $t$,  we use the
fifth-order phase integral approximation and compare this results
with the slow-roll and uniform approximation.

\subsection{Phase-integral approximation }
\begin{figure}[htbp]
\begin{center}
\subfigure[]{
\label{ch2:QSa}
\includegraphics[scale=0.4]{ch2_QS.eps}}
\subfigure[]{
\label{ch2:QSb}
\includegraphics[scale=0.35]{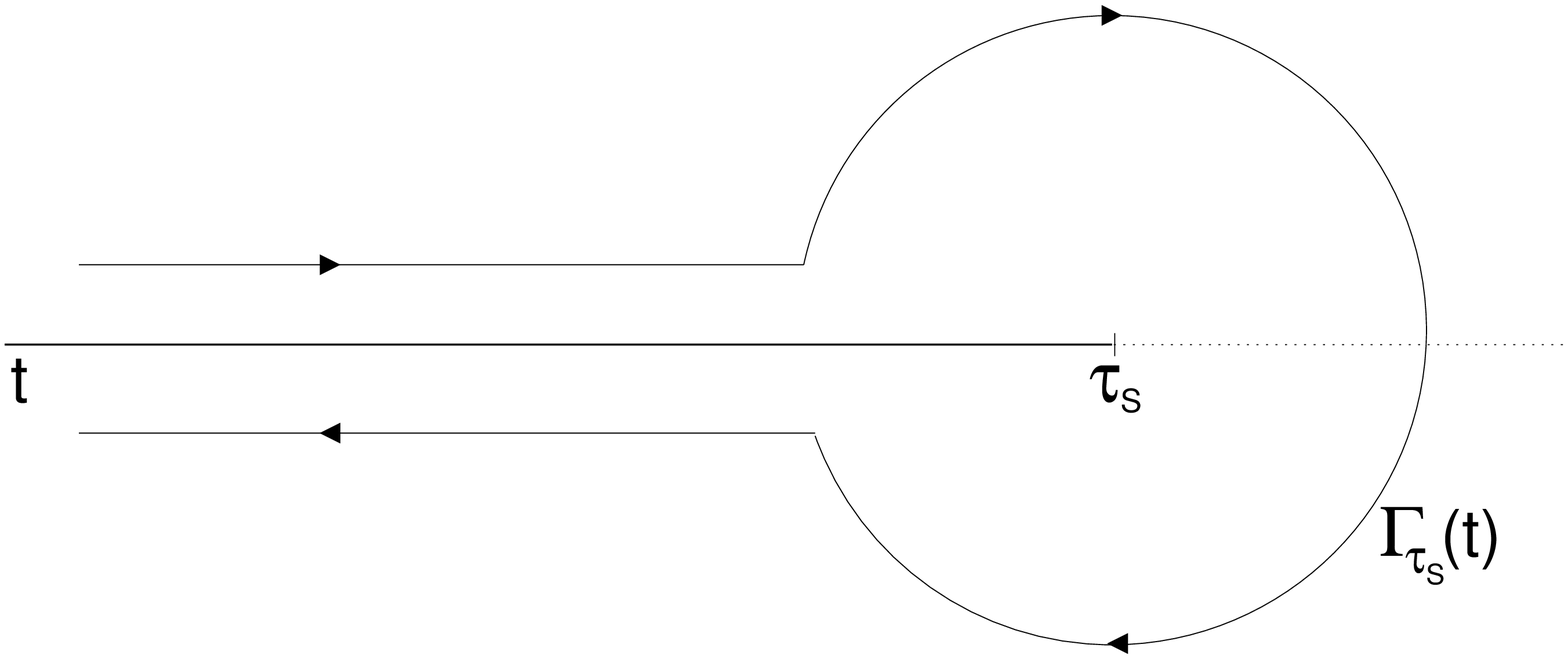}}
\subfigure[]{
\label{ch2:QSc}
\includegraphics[scale=0.35]{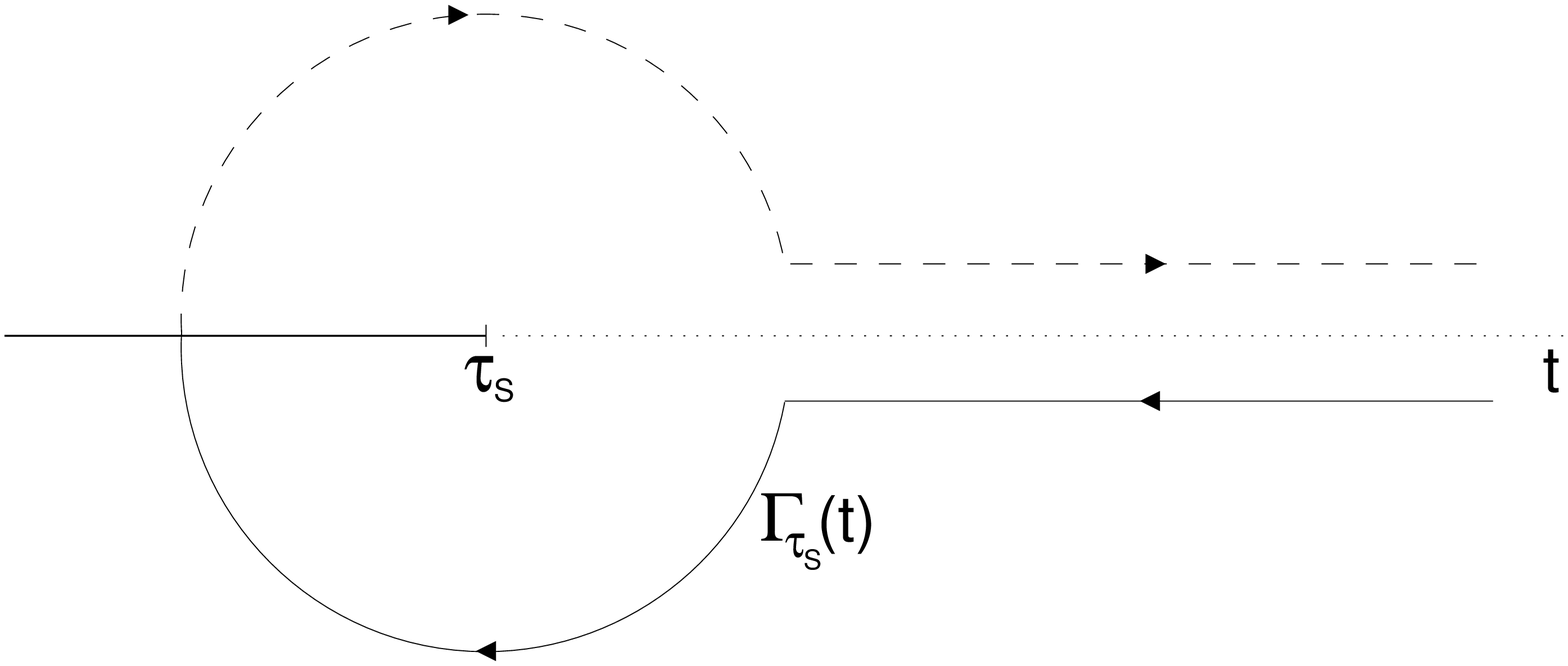}}
\caption{\small{(a) Behavior of the function  $Q_\sca^2(k,t)$. (b) Contour of integration $\Gamma_{\tau_\sca}(t)$ for  $0<t<\tau_\sca$. (c) Contour of  integration  $\Gamma_{\tau_\sca}(t)$ for $t>\tau_\sca$. The dashed line indicates the part of the path on the second Riemann sheet.}}
\end{center}
\end{figure}

\begin{figure}[htbp]
\begin{center}
\subfigure[]{
\label{ch2:QTa}
\includegraphics[scale=0.4]{ch2_QT.eps}}
\subfigure[]{
\label{ch2:QTb}
\includegraphics[scale=0.35]{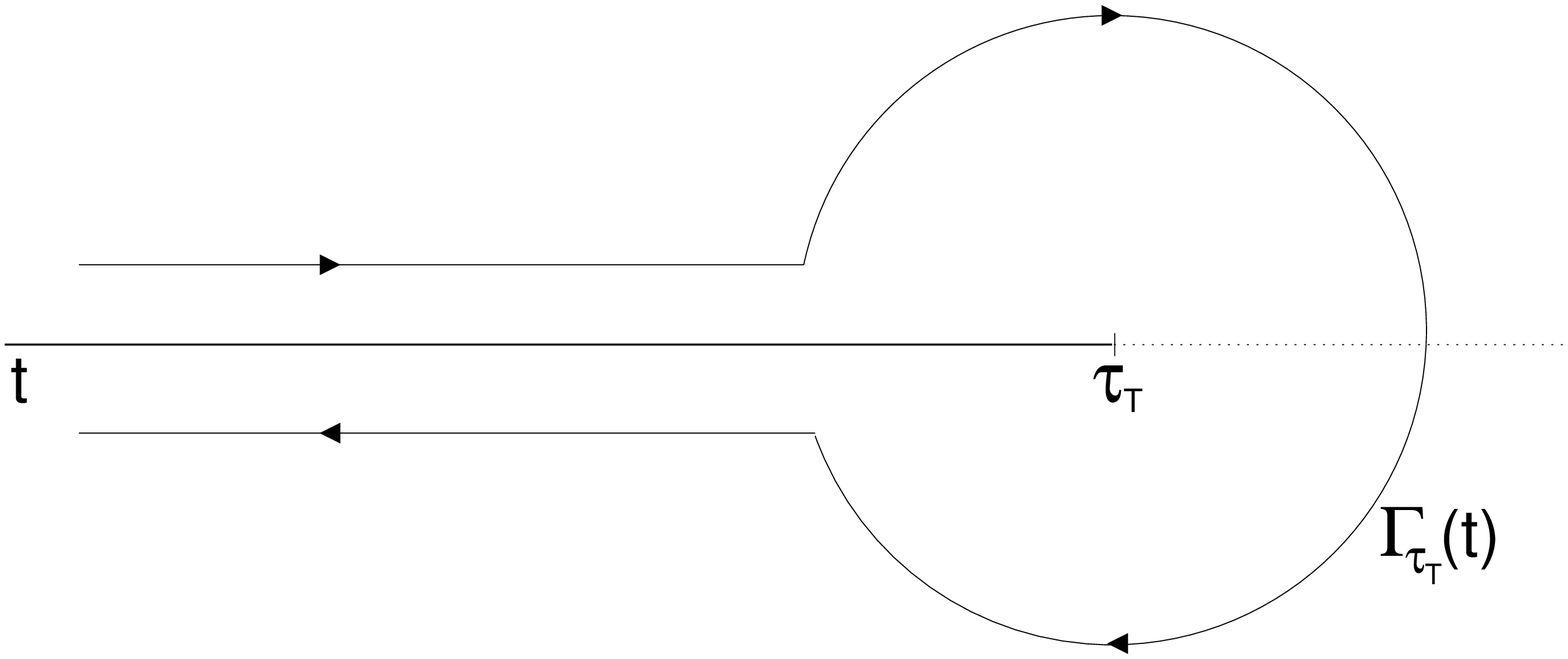}}
\subfigure[]{
\label{ch2:QTc}
\includegraphics[scale=0.35]{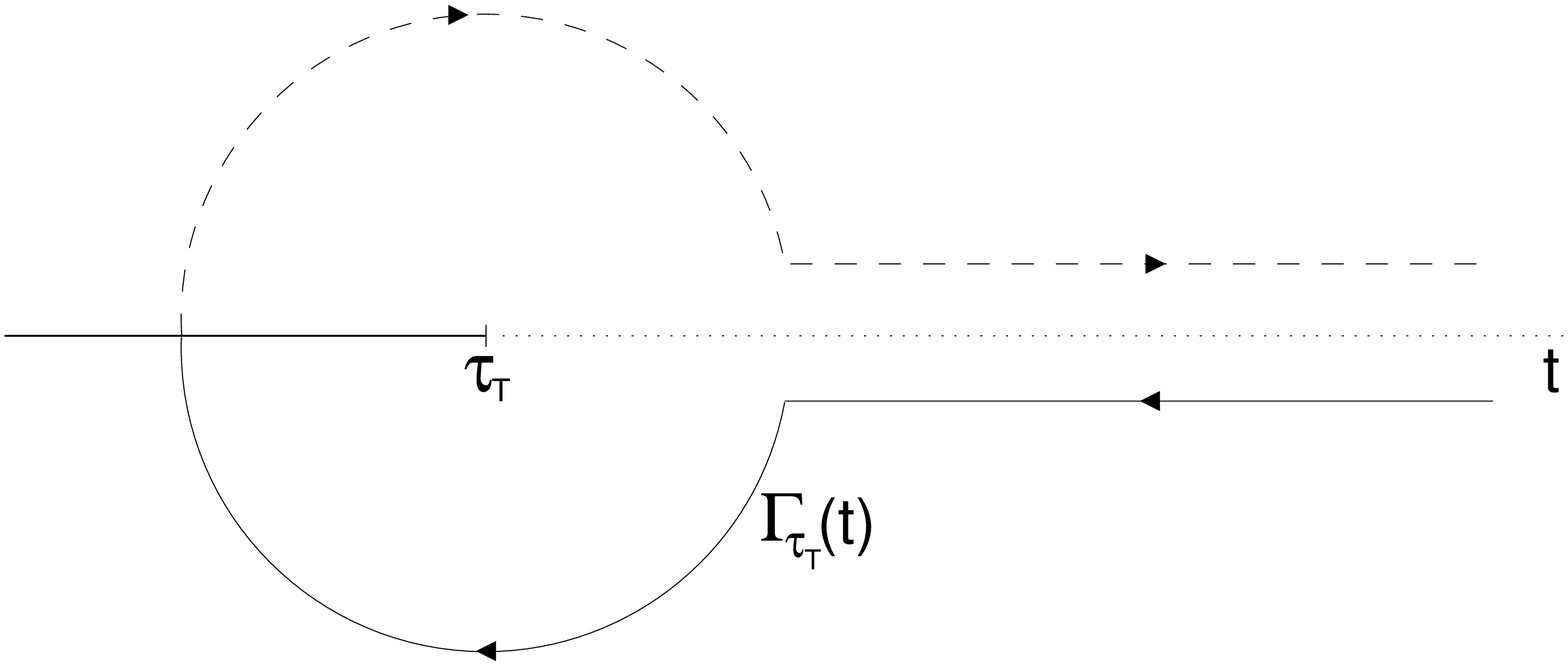}}
\caption{\small{(a) Behavior of  $Q_\ten^2(k,t)$.
(b) Contour of integration $\Gamma_{\tau_\ten}(t)$ for $0<t<\tau_\ten$.
(c) Contour of integration  $\Gamma_{\tau_\ten}(t)$ for $t>\tau_\ten$.
The dashed lined indicates the part of the path on the second Riemann sheet.}}
\end{center}
\end{figure}

In order to solve Eq. \eqref{ch2_ddotUk} and  Eq. \eqref{ch2_ddotVk}
with the help of the phase-integral approximation, we choose the
following base functions $Q$ for the scalar and tensor perturbations

\begin{eqnarray}
\label{Q}
Q_\sca^2(k,t)&=&R_\sca(k,t),\\
Q_\ten^2(k,t)&=&R_\ten(k,t),
\end{eqnarray}
where  $R_\sca(k,t)$ and  $R_\ten(k,t)$ are given by  Eq. \eqref{ch2_RS} and \eqref{ch2_RT} respectively.  Using this selection, the phase-integral approximation is  valid as  $k t\rightarrow \infty$, limit where we should impose the
condition \eqref{ch2_cero_Uk}, where the validity condition   $\mu \ll 1$ holds. The selection, given in Eq. (\ref{Q}), makes the first order phase-integral approximation coincide with the WKB solution. The bases functions  $Q_\sca(k,t)$ and  $Q_\ten(k,t)$ possess turning points  $t_\ret=\tau_\sca=1.38096\times 10^6 \,M_\Pl^{-1}$
and $t_\ret=\tau_\ten=t=1.38196\times 10^6 \,M_\Pl^{-1}$,  respectively for the mode $k=1.369\,h\,\Mpc^{-1}$. The turning point represents the horizon.  There are two ranges  where to define the solution. To the left of the turning point  $0<t<t_\ret$ we have the classically permitted region  $Q_{\sca,\ten}^2(k,t)>0$ and to the right of the turning point $t>t_\ret$ corresponding to the classically forbidden region
$Q_{\sca,\ten}^2(k,t)<0$, such as it is shown in Figs \ref{ch2:QSa}  and Fig. \ref{ch2:QTa}.

The mode  $k$ equations for the scalar an tensor perturbations
(\ref{ch2_ddotUk})  and  (\ref{ch2_ddotVk})in the phase-integral
approximation has two solutions:  For $0<t <t_\ret$

\begin{eqnarray}
\label{ch2_uk_left}
u^\phai_k(t)&=& \frac{c_1}{\sqrt{a(t)}}\left|q_\sca^{-1/2}(k,t)\right| \cos{\left[\left|\omega_\sca(k,t)\right|-\frac{\pi}{4}\right]} \\
\nonumber
&+& \frac{c_2}{\sqrt{a(t)}}\left|q_\sca^{-1/2}(k,t)\right| \cos{\left[\left|\omega_\sca(k,t)\right|+\frac{\pi}{4}\right]},\\
\label{ch2_vk_left}
v^\phai_k(t)&=& \frac{d_1}{\sqrt{a(t)}}\left|q_\ten^{-1/2}(k,t)\right| \cos{\left[\left|\omega_\ten(k,t)\right|-\frac{\pi}{4}\right]} \\
\nonumber
&+ &\frac{d_2}{\sqrt{a(t)}} \left|q_\ten^{-1/2}(k,t)\right| \cos{\left[\left|\omega_\ten(zk,t\right|+\frac{\pi}{4}\right]}.
\end{eqnarray}

and for $t>t_\ret$

\begin{eqnarray}
\label{ch2_uk_right}
u^\phai_k(t)&=&\frac{c_1}{2\sqrt{a(t)}}\left|q_\sca^{-1/2}(k,t)\right|\exp\left[-\left|\omega_\sca(k,t)\right|\right]\\
 \nonumber
&+& \frac{c_2}{\sqrt{a(t)}} \left|q_\sca^{-1/2}(k,t)\right| \exp\left[\left|\omega_\sca(k,t)\right|\right],\\
\label{ch2_vk_right}
v^\phai_k(z)&=&\frac{d_1}{2\sqrt{a(t)}}\left|q_\ten^{-1/2}(k,t)\right|\exp\left[-\left|\omega_\ten(k,t)\right|\right]\\
 \nonumber
& +& \frac{d_2}{\sqrt{a(t)}} \left|q_\ten^{-1/2}(k,t)\right| \exp\left[\left|\omega_\ten(k,t)\right|\right].
\end{eqnarray}

Using the phase-integral approximation up to fifth order ($2N+1=5\rightarrow N=2$), we have that $q_{\sca}(k,t)$ and  $q_{\ten}(k,t)$ can be expanded in the form

\begin{eqnarray}
\label{q1}
q_\sca(k,t)=\sum_{n=0}^2 Y_{2n_\sca}(k,t) Q_\sca(k,t)=\left[Y_{0_\sca}(k,t)+Y_{2_\sca}(k,t)+Y_{4_\sca}(k,t)\right] Q_\sca(k,t),\\
\label{q2}
q_\ten(k,t)=\sum_{n=0}^2 Y_{2n_\ten}(k,t)Q_\ten(k,t)=\left[Y_{0_\ten}(k,t)+Y_{2_\ten}(k,t)+Y_{4_\ten}(k,t)\right] Q_\ten(k,t).
\end{eqnarray}
In order to compute $q_\sca(k,t)$ and $q_\ten(k,t)$, we  compute $Y_{2_\sca}(k,t)$, $Y_{4_\sca}(k,t)$, $Y_{2_\ten}(k,t)$, $Y_{4_\ten}(k,t)$ and the required functions  $\varepsilon_{0_\sca}(k,t)$, $\varepsilon_{2_\sca}(k,t)$, $\varepsilon_{0_\ten}(k,t)$ and $\varepsilon_{2_\ten}(k,t)$.  The expressions  (\ref{q1}) and (\ref{q2}) give  a fifth-order approximation for $q_\sca(k,t)$  and $q_\ten(k,t)$.  In order to compute $\omega_\sca(k,t)$ and $\omega_\ten(k,t)$ we make a contour integration following the path indicated in Figs. \ref{ch2:QSb}-(c) \ref{ch2:QTb}-(c)

\begin{eqnarray}
\omega_\sca(k,t)&=&\omega_{0_\sca}(k,t)+\sum_{n=1}^{2} \omega_{2n_\sca}(k,t),\\
&=&\int_{\tau_\sca}^{t}Q_\sca(k,t)\D t+\frac{1}{2}\sum_{n=1}^2\int_{\Gamma_{\tau_\sca}}Y_{2n_\sca}(k,t)Q_\sca(k,t)\D t,\\
&=&\int_{\tau_\sca}^{t}Q_\sca(k,t)\D t+\frac{1}{2}\sum_{n=1}^2\int_{\Gamma_{\tau_\sca}}f_{2n_\sca}(k,t)\D t,\\
\omega_\ten(k,t)&=&\omega_{0_\ten}(k,t)+\sum_{n=1}^{2} \omega_{2n_\ten}(k,t),\\
&=&\int_{\tau_\ten}^{t}Q_\ten(k,t)\D t+\frac{1}{2}\sum_{n=1}^2\int_{\Gamma_{\tau_\ten}}Y_{2n_\ten}(k,t)Q_\ten(k,t)\D t,\\
&=&\int_{\tau_\sca}^{t}Q_\ten(k,t)\D z+\frac{1}{2}\sum_{n=1}^2\int_{\Gamma_{\tau_\ten}}f_{2n_\ten}(k,t)\D t,
\end{eqnarray}
\medskip
where
\begin{eqnarray}
f_{2n_\sca}(k,t)&=&Y_{2n_\sca}(k,t)Q_\sca(k,t),\\
f_{2n_\ten}(k,t)&=&Y_{2n_\ten}(k,t)Q_\ten(k,t).
\end{eqnarray}
The functions  $f_{2n_\sca}(k,t)$ and  $f_{2n_\ten}(k,t)$ have
the following functional dependence:

\begin{eqnarray}
\label{A}
f_{2_\sca}(k,t)&=&A(k,t)(t-\tau_\sca)^{-5/2},\\
f_{4_\sca}(k,t)&=&B(k,t)(t-\tau_\sca)^{-11/2},\\
f_{2_\ten}(k,t)&=&C(k,t)(t-\tau_\ten)^{-5/2},\\
\label{D}
f_{4_\ten}(k,t)&=&D(k,t)(t-\tau_\ten)^{-11/2},
\end{eqnarray}
where the functions $A(k,t)$ and $B(k,t)$ are regular at $\tau_\sca$
and the functions $C(k,t)$, $D(k,t)$ are regular at  $\tau_\ten$.
With the help of  the functions  \eqref{A}-\eqref{D} we compute the
integrals for $\omega_{2n}$ up to $N=4$ using the contour indicated
in  Figs. \ref{ch2:QSb}-(c) and \ref{ch2:QTb}-(c). The expressions
for  $\omega_{2n}$ permit one to obtain the fifth-order phase
integral approximation of the solution to the equations for scalar
\eqref{ch2_ddotUk} and tensor \eqref{ch2_ddotVk} perturbations.  The
constants $c_1$, $c_2$, $d_1$ and  $d_2$ are obtained using the
limit  $k\,t\rightarrow 0$ of the solutions on the left side of the
turning point \eqref{ch2_uk_left} and \eqref{ch2_vk_left}, and  are
given by the expressions

\begin{eqnarray}
c_1&=&-\im\,c_2,\\
c_2&=&\frac{\e^{-\im\frac{\pi}{4}}}{\sqrt{2}}\e^{-\im\left[k\,\eta(0)+\left|\omega_{0_\sca}(k,0)\right|\right]},\\
d_1&=&-\im\,d_2,\\
d_2&=&\frac{\e^{-\im\frac{\pi}{4}}}{\sqrt{2}}\e^{-\im\left[k\,\eta(0)+\left|\omega_{0_\ten}(k,0)\right|\right]},
\end{eqnarray}
In order to compute the scalar and tensor power spectra, we need to
calculate the limit as $k\,t\rightarrow \infty$ of the growing part
of the solutions on the right side of the turning point  given by
Eq. \eqref{ch2_uk_right} and  Eq. \eqref{ch2_vk_right} for scalar
and tensor perturbations respectively.

\begin{eqnarray}
P_\sca(k)&=&\lim_{-k t\rightarrow \infty} \frac{k^3}{2\pi^2} \left|\frac{u_k^\phai(t)}{z_\sca(t)}\right|^2,\\
P_\ten(k)&=&\lim_{-k t\rightarrow \infty} \frac{k^3}{2\pi^2} \left|\frac{v_k^\phai(t)}{a(t)}\right|^2.
\end{eqnarray}

\subsection{Uniform approximation}
We want to obtain an approximate solution to the differential
equations \eqref{ch2_ddotUk} and \eqref{ch2_ddotVk} in the range
where  $Q_\sca^2(k,t)$ and   $Q_\ten^2(k,t)$ have a simple root  at
$t_\ret=\tau_\sca$, and $t_\ret=\tau_\ten$, respectively, so that
$Q_{\sca,\ten}^2(k,t)>0$ for  $0<t<t_\ret$ and
$Q_{\sca,\ten}^2(k,t)<0$ for  $t>t_\ret$ as depicted in Fig.
\ref{ch2:QSa} and Fig. \ref{ch2:QTa}. Using the uniform
approximation method
\cite{Berry,habib:2002,rojas:2007b,rojas:2007c}, we obtain that for
$0<t<t_\ret$ we have

\begin{eqnarray}
\label{Uk_zero}
U_k(k,t)&=&\left[\frac{\rho_\ele(k,t)}{Q_\sca^2(k,t)} \right]^{1/4} \left\{C_1
A_i[-\rho_\ele(k,t)]+C_2 B_i[-\rho_\ele(k,t)] \right\},\\
\label{Vk_zero}
V_k(k,t)&=&\left[\frac{\rho_\ele(k,t)}{Q_\ten^2(k,t)} \right]^{1/4} \left\{C_1
A_i[-\rho_\ele(k,t)]+C_2 B_i[-\rho_\ele(k,t)] \right\},\\
\frac{2}{3}\left[\rho_\ele(k,t)\right]^{3/2}&=&\int_{t}^{t_\ret} \left[Q_{\sca,\ten}^2(k,t)\right]^{1/2}\D t,
\end{eqnarray}
\medskip
where  $C_1$ and  $C_2$ are two constants to be determined with the help of the boundary conditions \eqref{ch2_borde_Uk}. For $t> t_\ret$

\begin{eqnarray}
\label{Uk_infinity}
U_k(k,t)&=&\left[\frac{-\rho_\ere(k,t)}{Q_\sca^2(k,t)} \right]^{1/4} \left\{C_1
A_i[\rho_\ere(k,t)]+C_2 B_i[\rho_\ere(k,t)] \right\},\\
\label{Vk_infinity}
V_k(k,t)&=&\left[\frac{-\rho_\ere(k,t)}{Q_\ten^2(k,t)} \right]^{1/4} \left\{C_1
A_i[\rho_\ere(k,t)]+C_2 B_i[\rho_\ere(k,t)] \right\},\\
\frac{2}{3}\left[\rho_\ere(k,t)\right]^{3/2}&=&\int_{t_\ret}^{t} \left[-Q_{\sca,\ten}^2(k,t)\right]^{1/2}\D t,
\end{eqnarray}

For the computation of the power spectrum we need to take the limit
$k\,t\rightarrow \infty$ of the solutions  \eqref{Uk_infinity} and
\eqref{Vk_infinity}. In this limit we have

\begin{eqnarray}
\label{limit_uk}
u_k^\ua(t)&\rightarrow&  \frac{C}{\sqrt{2\,a(t)}}\left[-Q_\sca^2(k,t)\right]^{-1/2}\left\{ \frac{1}{2}\exp\left(-\int_{\tau_\sca}^{t}\left[-Q_\sca^2(k,t)\right]^{1/2} \D t\right)\right.\\
\nonumber
& +&\left.\im\,\exp\left(\int_{\tau_\sca}^{t}\left[-Q_\sca^2(k,t)\right]^{1/2} \D t\right)\right\}\\
\label{limit_vk}
v_k^\ua(t)&\rightarrow&  \frac{C}{\sqrt{2\,a(t)}}\left[-Q_\ten^2(k,t)\right]^{-1/2}\left\{ \frac{1}{2}\exp\left(-\int_{\tau_\ten}^{t}\left[-Q_\ten^2(k,t)\right]^{1/2} \D t\right)\right.\\
\nonumber
& +&\left.\im\,\exp\left(\int_{\tau_\ten}^{t}\left[-Q_\ten^2(k,t)\right]^{1/2} \D t\right)\right\},
\end{eqnarray}
where $C$ is a phase factor. Notice that  Eq.  \eqref{limit_uk} and Eq. \eqref{limit_vk} are identical to Eq.  \eqref{ch2_uk_right} and Eq. \eqref{ch2_vk_right}  obtained in the first-order phase-integral approximation.
Using Eqs. \eqref{ch2_a_fit},  \eqref{ch2_zs} and the growing part  of the solutions   \eqref{limit_uk} and  \eqref{limit_vk}   one can compute the scalar and tensor power spectrum using the uniform approximation method,

\begin{eqnarray}
P_\sca(k)&=&\lim_{-k t\rightarrow \infty} \frac{k^3}{2\pi^2} \left|\frac{u_k^\ua(t)}{z_\sca(t)}\right|^2,\\
P_\ten(k)&=&\lim_{-k t\rightarrow \infty} \frac{k^3}{2\pi^2} \left|\frac{v_k^\ua(t)}{a(t)}\right|^2.
\end{eqnarray}
We also use the second-order improved uniform approximation for the power spectrum \cite{habib:2005B},

\begin{equation}
 \tilde{P}_{\sca,\ten}(k)=P_{\sca,\ten}(k)\left[\Gamma^*(\bar{\nu}_{\sca,\ten})\right],
\end{equation}
where $\bar{\nu}_{\sca,\ten}$ is the turning point for the scalar or tensor power spectrum and

\begin{equation}
\Gamma^*(\nu)\equiv 1+\frac{1}{12\nu}+\frac{1}{288\nu}-\frac{139}{51840\nu}+\cdots.
\end{equation}

\subsection{Slow-roll approximation}
The scalar and tensor power spectra in the slow-roll approximation
to second-order are given by the expressions
\cite{stewart:2001,gong:2004}

 \begin{eqnarray}
\label{ch2_sr_PS}
P_\sca^{\sr}(k)&\simeq&\left[1+(4c-2)\epsilon_1+2c\delta_1+\left(3c^2+2c-22+\frac{29\pi^2}{12}\right)\epsilon_1\delta_1+\left(3c^2-4+\frac{5\pi^2}{12}\right)\delta_1^2+\right.\\
\nonumber
&+&\left.\left(-c^2+\frac{\pi^2}{12}\right)\delta_2\right]\left.\left(\frac{H}{2\pi}\right)^2\left(\frac{H}{\dot{\phi}}\right)^2\right|_{k=aH} ,\\
\label{ch2_sr_PT}
P_\ten^{\sr}(k)&\simeq& \left[1+(2c-2)\epsilon_1+\left(2c^2-2c-3+\frac{\pi^2}{2}\right)\epsilon_1^2+\left(-c^2+2c-2+\frac{\pi^2}{12}\right)\epsilon_2\right]\left.\left(\frac{H}{2\pi}\right)^2\right|_{k=aH},
\end{eqnarray}
where  $b$ is the Euler constant, $2-\ln2-b\simeq 0.7296$ and $\ln2+b-1\simeq 0.2704$, and

\begin{eqnarray}
\epsilon_1&=&\frac{\dot{H}}{H^2}=\frac{1}{1+2N_*},\\
\epsilon_2&=&\frac{1}{H}\frac{\D\epsilon_1}{\D t}=\frac{2}{(1+2N_*)^2},\\
\delta_n&\equiv&\frac{1}{H^n\dot{\phi}}\frac{\D^{n+1}\phi}{\D t^{n+1}}\rightarrow\delta_1=\delta_2=0.
\end{eqnarray}

The spectral index in the slow-roll approximation are

\begin{eqnarray}
\label{ch2_sr_nS}
n_\sca^{\sr}(k)&\simeq&1-4\epsilon_1-2\delta_1+(8c-8)\epsilon_1^2+(10c-6)\epsilon_1\delta_1 ,\\
\label{ch2_sr_nT}
n_\ten^{\sr}(k)&\simeq&-2\epsilon_1-2\epsilon_1^2+(2\alpha-2)\epsilon_2.
\end{eqnarray}

The expressions  \eqref{ch2_sr_PS}, \eqref{ch2_sr_PT},
\eqref{ch2_sr_nS} y \eqref{ch2_sr_nT}  depend explicitly on time. In
order to compute the scalar and tensor power spectra we need to
obtain the dependence  on the variable $k$. For a given value of $k$
($0.0001 \,\Mpc^{-1} \leq k \leq 15 \,\Mpc^{-1}$) we obtain  $t_*$
from the relation $k=aH$. Thus, for each  $k$ one obtains a value of
$t$ that  we substitute into $N_*$ and Eqs.  \eqref{ch2_sr_PS},
\eqref{ch2_sr_PT}, \eqref{ch2_sr_nS} y \eqref{ch2_sr_nT}.

\subsection{Numerical solution}
\begin{figure}[htbp]
\begin{center}
\subfigure[]{
\label{ch2:Re(uk)}
\includegraphics[scale=0.465]{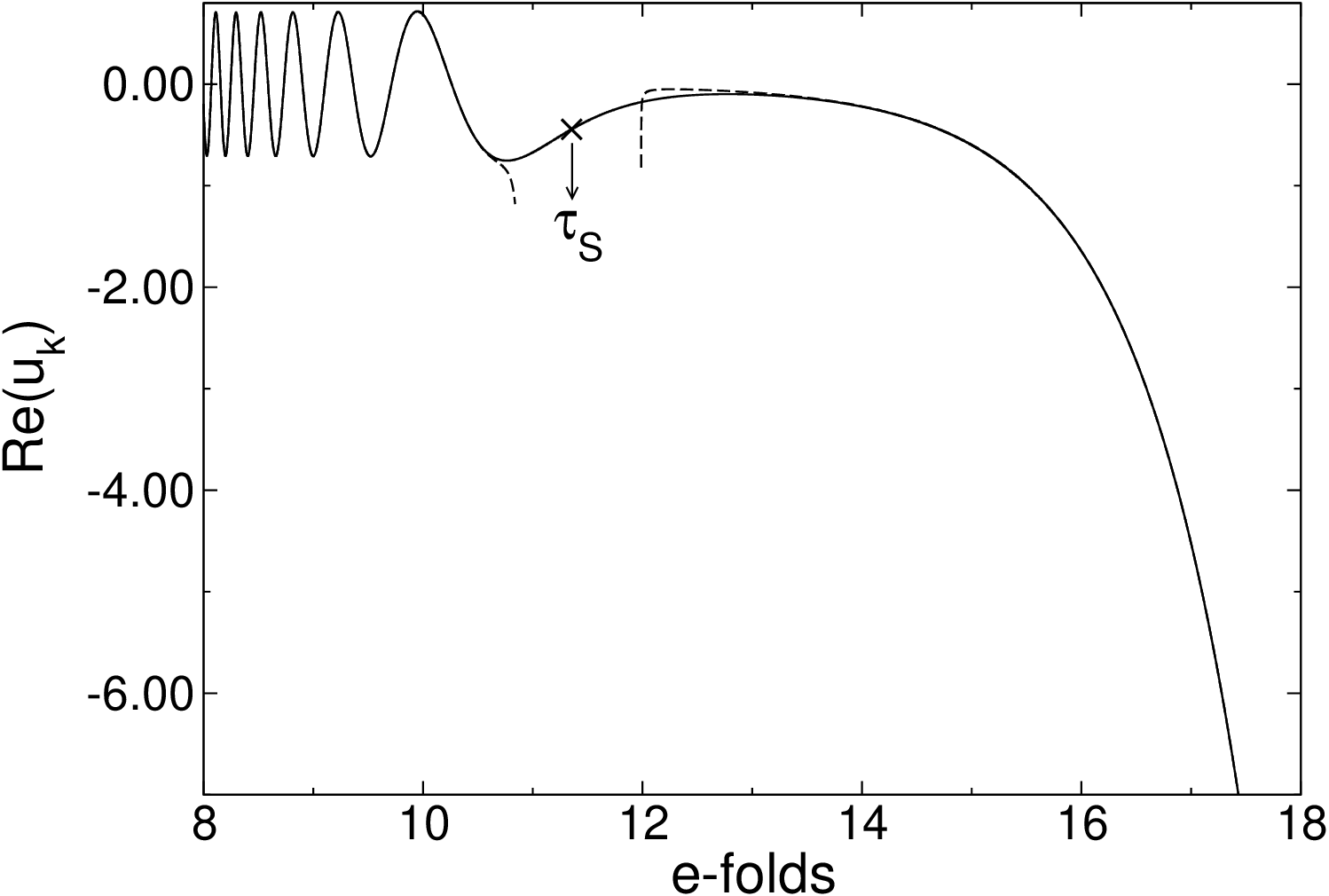}}\\
\vspace{1.2cm}
\subfigure[]{
\includegraphics[scale=0.465]{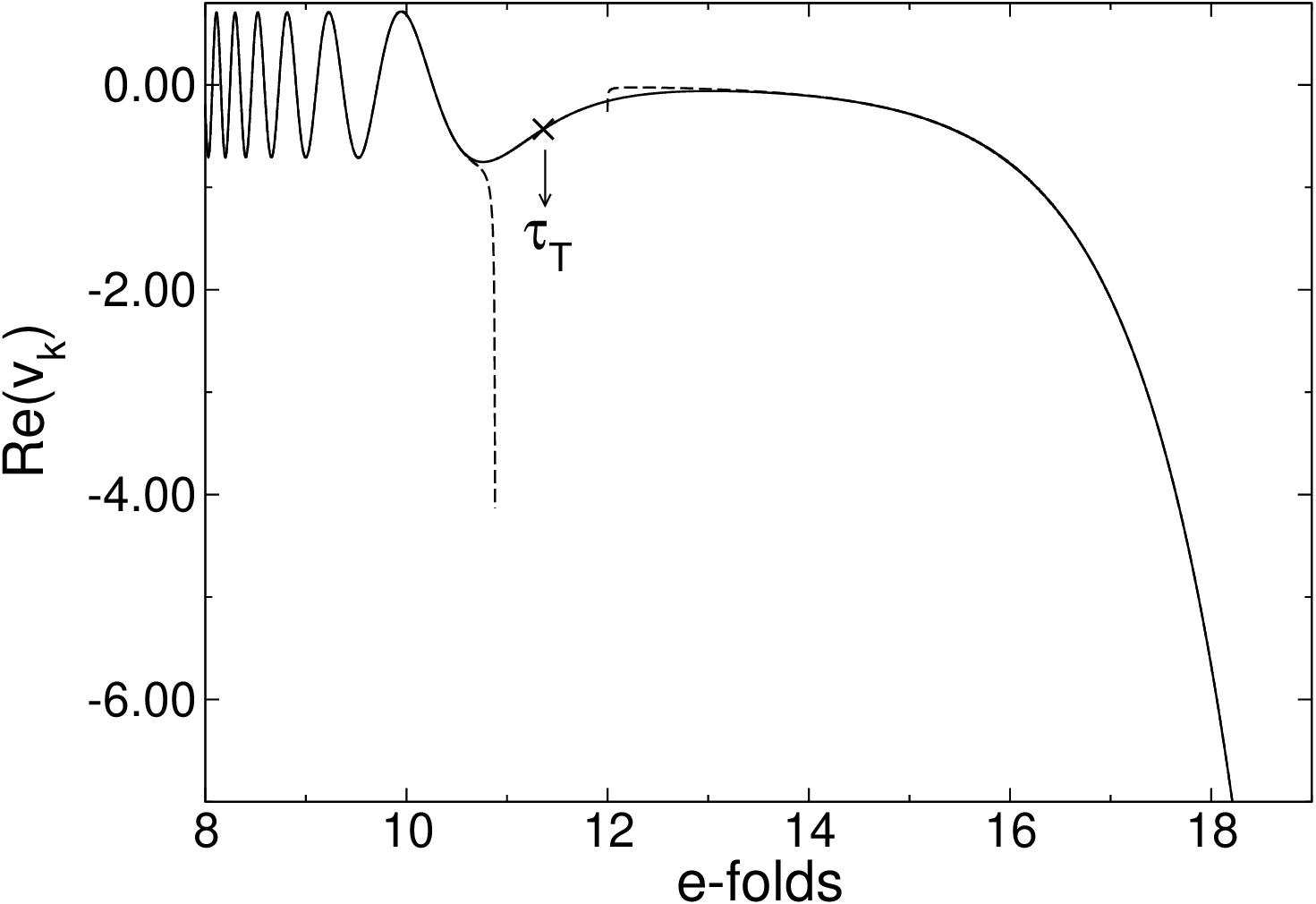}}
\caption{(a) $\real(u_k)$ and  (b) $\real(v_k)$  versus the number of  e-folds for the chaotic inflationary model  $\frac{1}{2}m^2\phi^2$.  Solid line: numerical solution (ICs phi3);  dashed line : fifth-order phase-integral approximation}
\end{center}
\end{figure}

\begin{figure}[htbp]
\begin{center}
\subfigure[]{\includegraphics[scale=0.465]{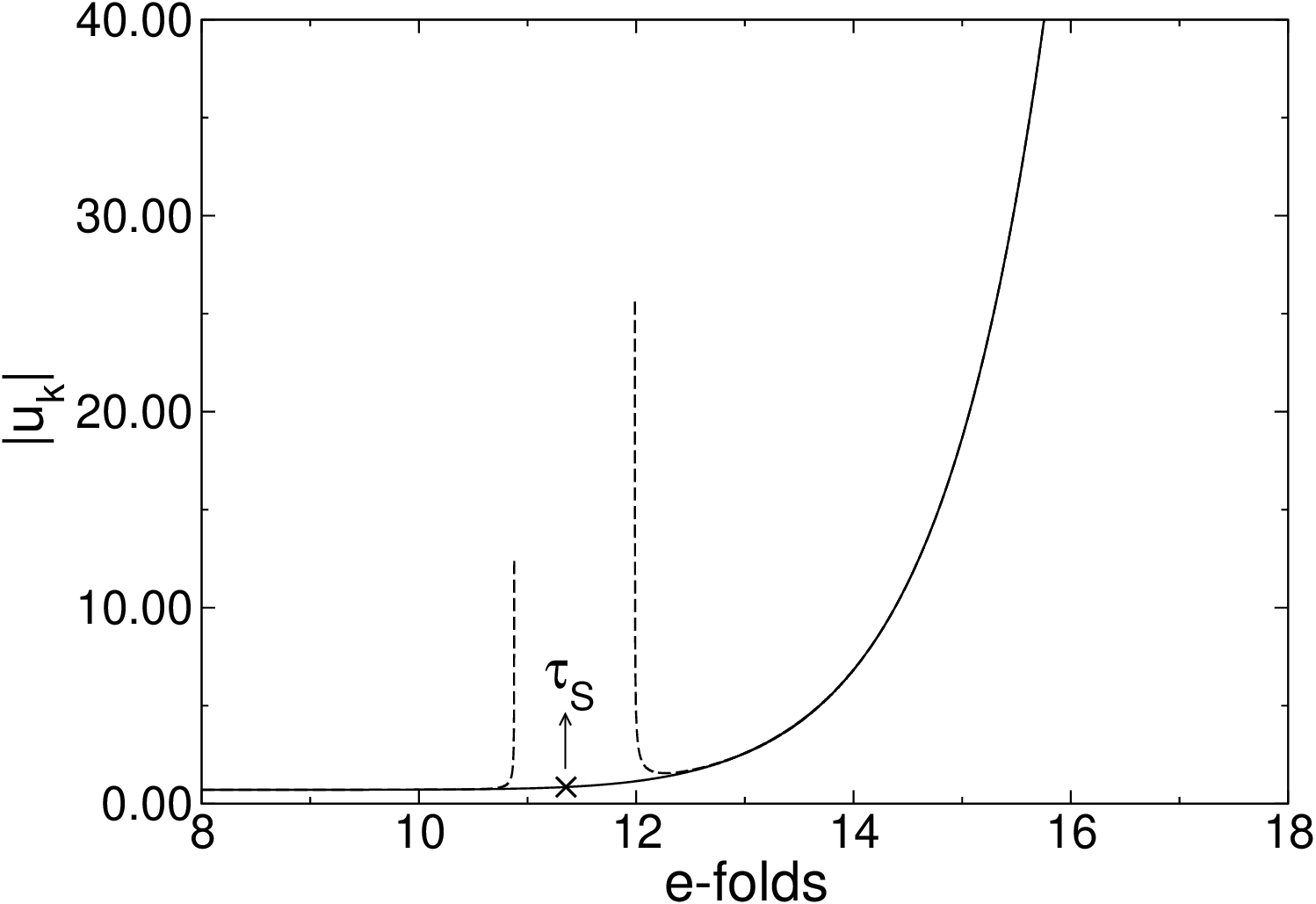}}\\
\vspace{1.2cm}
\subfigure[]{
\label{ch2:Abs(vk)}
\includegraphics[scale=0.465]{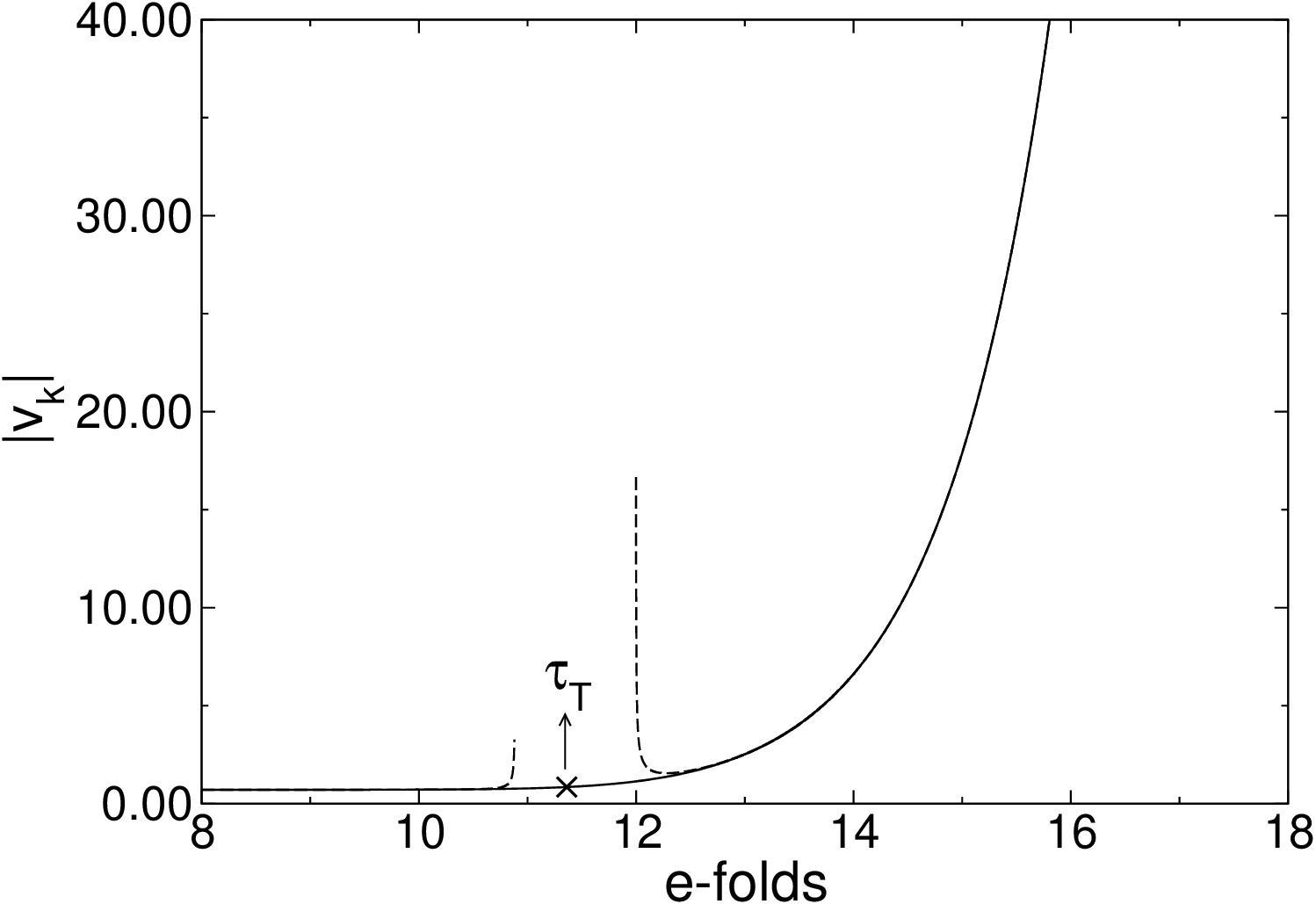}}
\caption{(a) $\left|u_k\right|$ and (b) $\left|v_k\right|$ versus the number of  de e-folds for the chaotic inflationary model $\frac{1}{2}m^2\phi^2$.  Solid line: Numerical result (ICs phi3);  dashed line: fifth-order phase-integral approximation}
\end{center}
\end{figure}

We integrate on the physical time $t$ the equations \eqref{ch2_ddotUk} and  \eqref{ch2_ddotVk}  governing the scalar and tensor perturbations  using the predictor-corrector Adams method of order $12$ \cite{gerald:1984}, and  solve two differential equations, one for the real part and another for the imaginary  part $U_k$ and $V_k$.  Two initial conditions are needed in each case  $U_k(t_i)$,  $U'_k(t_i)$, $V_k(t_i)$,  $V'_k(t_i)$, which can be obtained from the third-order
phase-integral approximation.  We start the numerical integration at $t_i$ calculated at $25$ oscillations before  reaching the turning  point $t_\ret$ \cite{cunha:2005}. We call this procedure  ICs phi3. Figs. \ref{ch2:Re(uk)}-\ref{ch2:Abs(vk)}  compare the numerical solution with the fifth-order phase-integral approximation  for
$\real(u_k)$, $\imag(u_k)$, $\left|u_k\right|$, $\real(v_k)$, $\imag(v_k)$,  and $\left|v_k\right|$. Figures are plotted against the number of e-folds   $N$.  The solid line corresponds to the numerical solution (ICs phi3), the dashed line  corresponds to the fifth-order phase-integral approximation.  In each case the turning
point $\tau_\sca$ and $\tau_\ten$ are indicated with an arrow. We stop the numerical computation of  $P_\sca(k)$ and  $P_\ten(k)$ at $t= 5.00\times10^6\,M_\Pl^{-1}$, after  the mode leaves the horizon, where  $u_k/z_\sca$ and  $v_k/a$ are approximately constant. Notice that the expressions for  fitting  \eqref{ch2_a_fit} and \eqref{ch2_phi_fit} are valid in the aforementioned  time scales, therefore we can
use them for computing  the scalar $P_\sca(k)$ and tensor $P_\ten(k)$  power spectra.

\begin{figure}[htbp]
\begin{center}
\subfigure[]{
\includegraphics[scale=0.465]{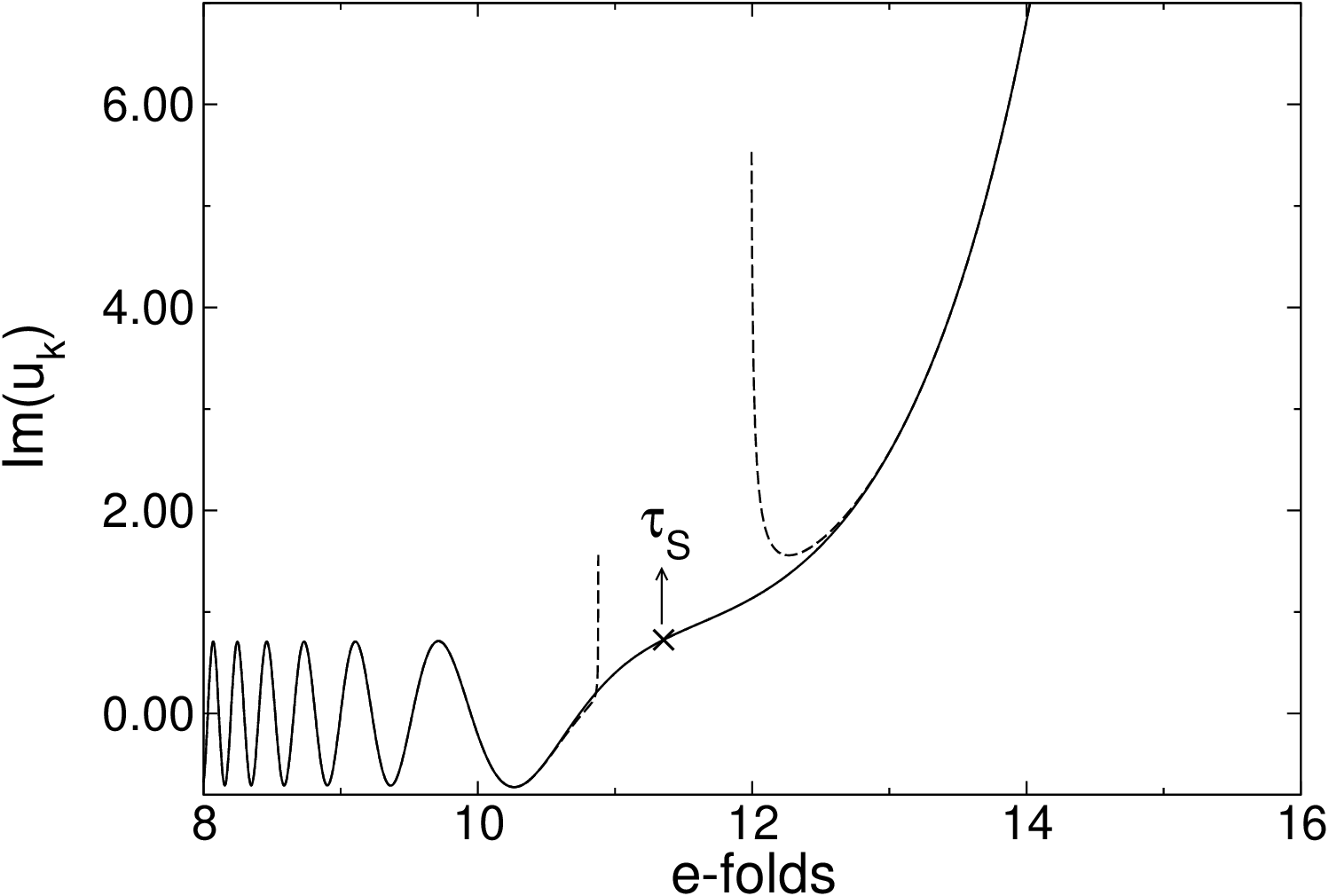}}\\
\vspace{1.2cm}
\subfigure[]{\includegraphics[scale=0.465]{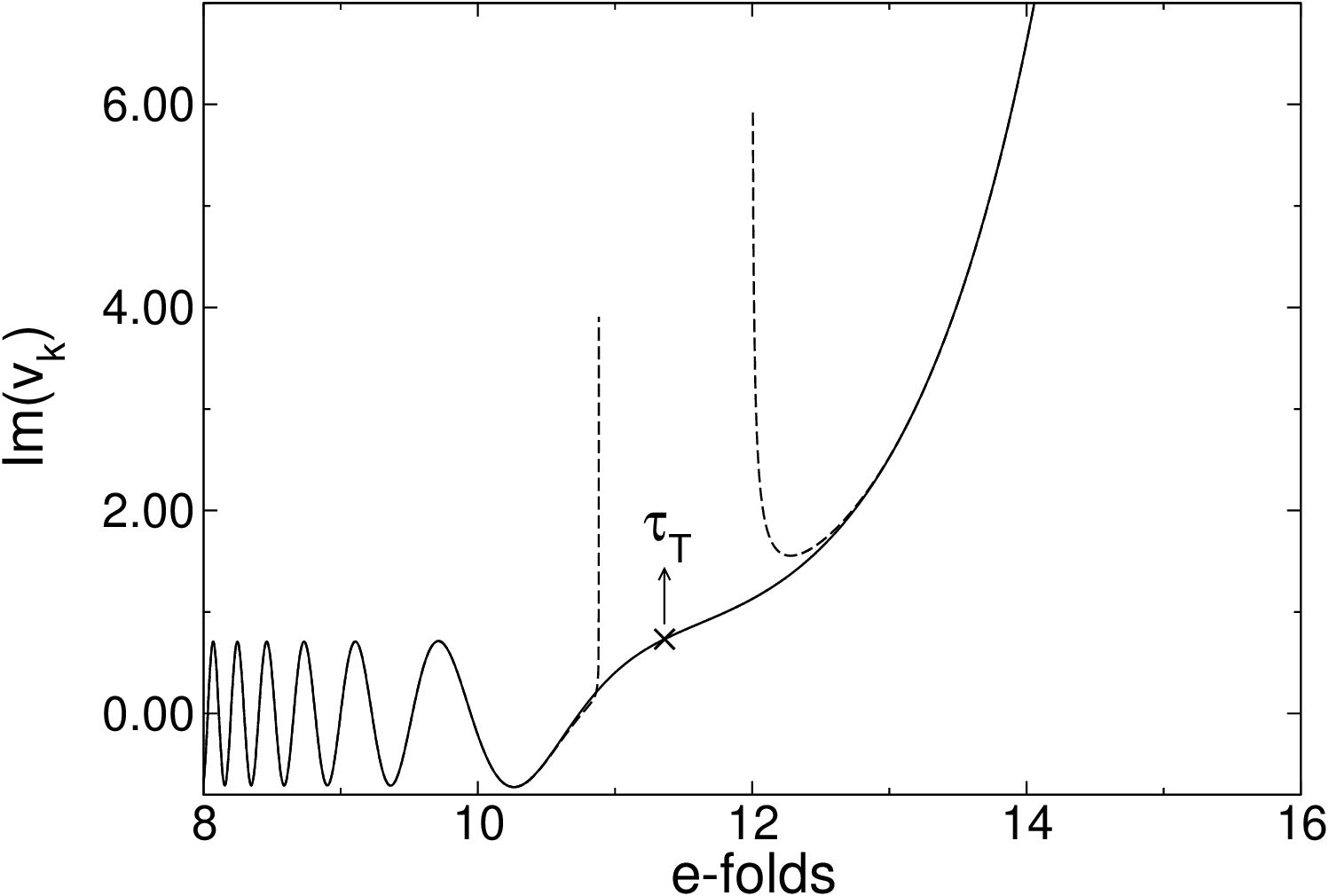}}
\caption{(a) $\imag(u_k)$ and (b) $\imag(v_k)$  versus the number of e-folds for the chaotic inflationary model $\frac{1}{2}m^2\phi^2$. Solid line : numerical result  (ICs phi3);  dashed line: fifth-order phase-integral approximation}
\end{center}
\end{figure}

\subsection{Results}\label{ch2:results}
\begin{figure}[thbp]
\begin{center}
\includegraphics[scale=0.465]{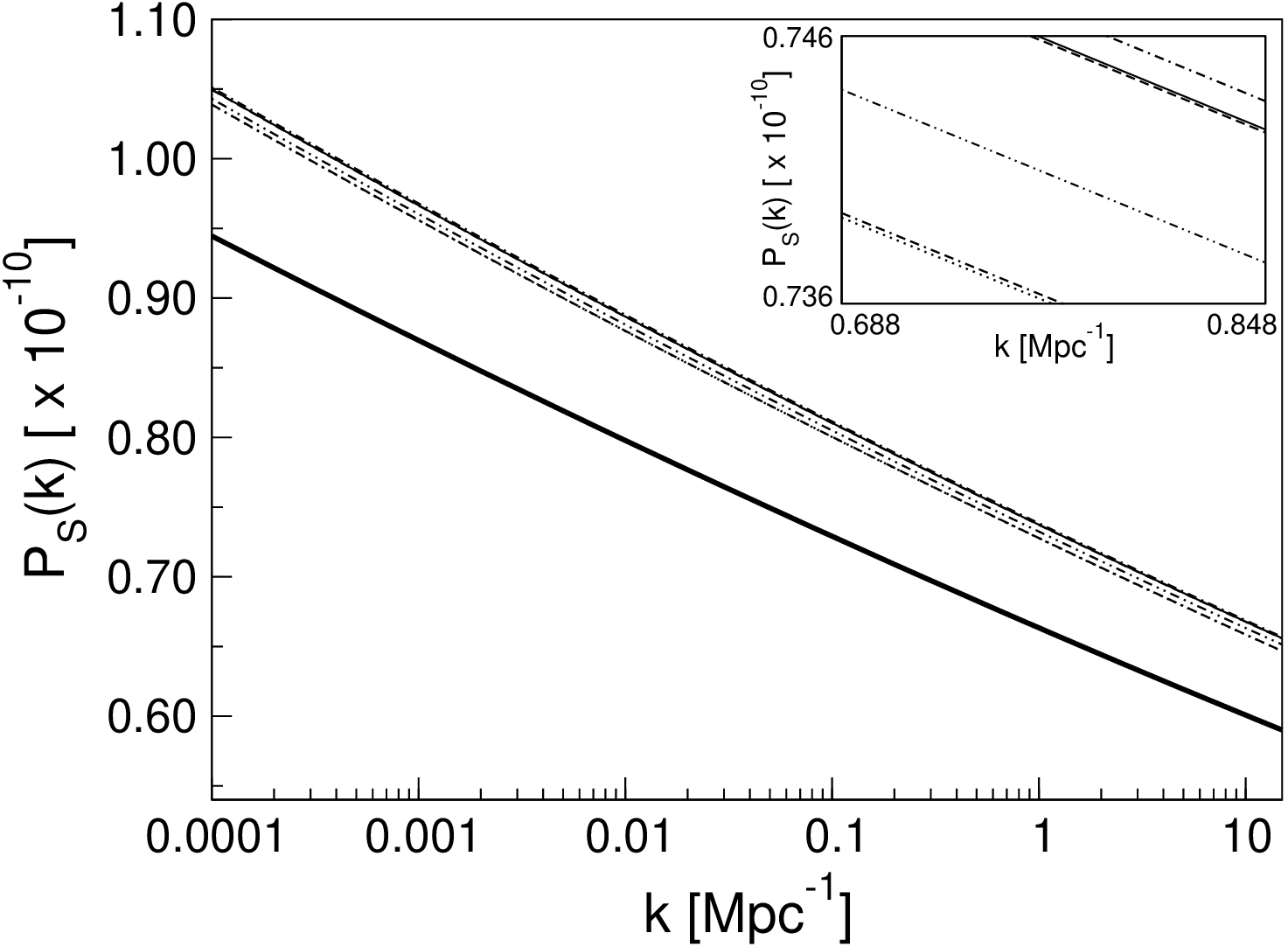}
\caption{\small{$P_\sca(k)$  for the chaotic inflationary $\frac{1}{2}m^2\phi^2$ model.  Thin solid line: numerical result (ICs phi3); dot-dashed line: third-order phase-integral approximation, dashed line: fifth-order phase-integral approximation; thick solid line: first-order phase-integral approximation, WKB and  first-order uniform approximation, dashed double-dots line: second-order improved uniform approximation, double-dashed dot line: second-order slow-roll approximation, dotted line: first-order slow-roll approximation. The inset is an enlargement of the figure.}}
\label{ch2:PS}
\end{center}
\end{figure}

\begin{figure}[thbp]
\begin{center}
\includegraphics[scale=0.465]{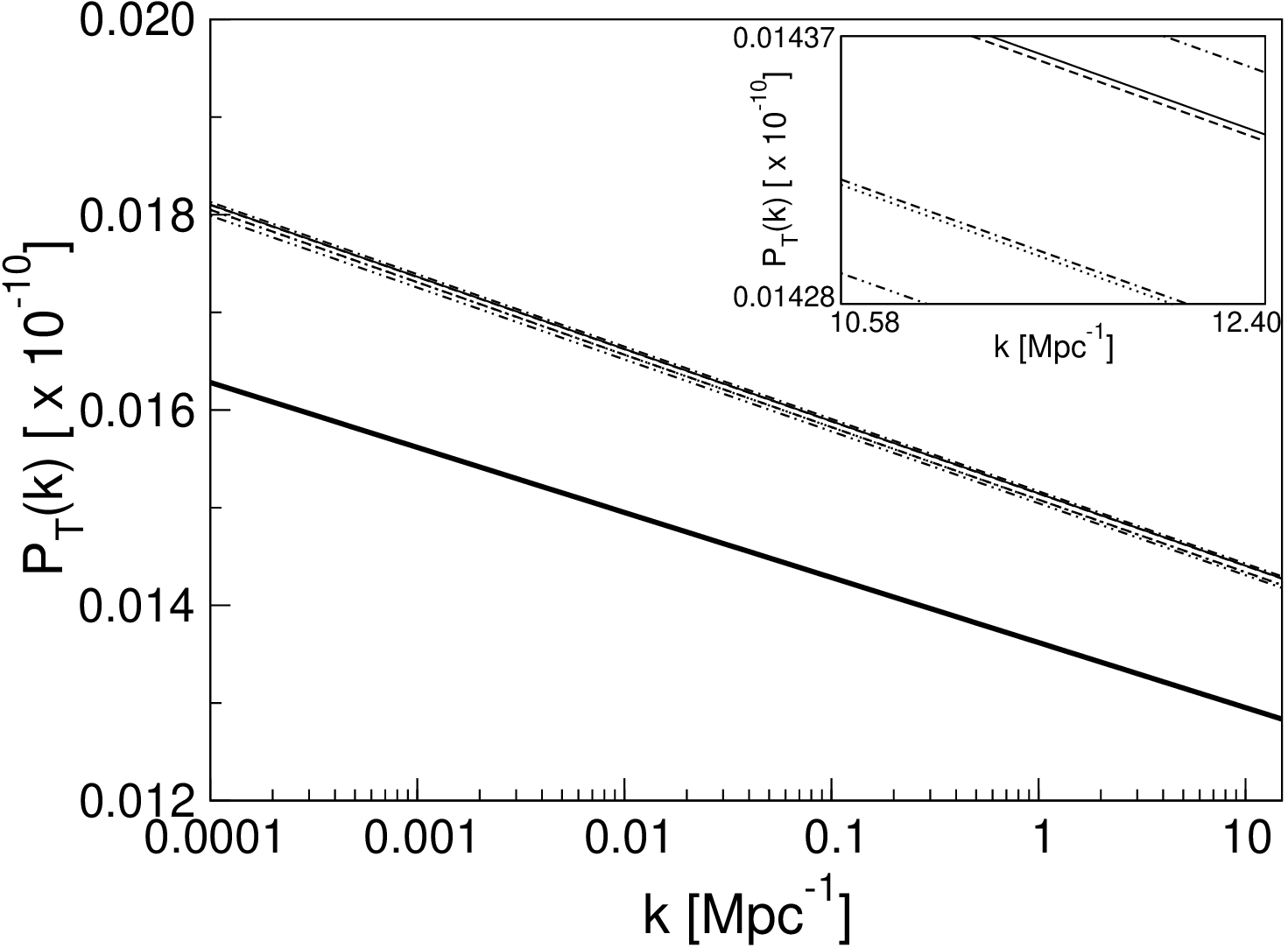}
\caption{\small{$P_\ten(k)$  for the chaotic inflationary $\frac{1}{2}m^2\phi^2$ model.  Thin solid line: numerical result (ICs phi3); dot-dashed line: third-order phase-integral approximation; dashed line: fifth-order phase-integral approximation; thick solid line: first-order phase-integral approximation, WKB and  first-order uniform approximation; dashed double-dots line: second-order improved uniform approximation; double-dashed dot line: second-order slow-roll approximation; dotted line: first-order slow-roll approximation. The inset is an enlargement of the figure.}}
\label{ch2:PT}
\end{center}
\end{figure}

For the chaotic $\frac{1}{2}m^2\phi^2$ inflationary model,  we want to compare the scalar and tensor  power spectra and the spectral indices for different values of  $k$ calculated using the third and fifth-order  phase-integral approximation  with the numerical result (ICs phi3), the first and second-order slow-roll approximation and  the first and second-order uniform approximation method.  First we analyze the results for the scalar $P_\sca(k)$ and tensor $P_\ten(k)$ power spectra   shown in Fig. \ref{ch2:PS} and Fig. \ref{ch2:PT}.

Table \ref{ch2_t1} shows the value of $P_\sca(k)$, $P_\ten(k)$, $n_\sca(k)$, and $n_\ten(k)$ using each method of approximation at the WMAP pivot scale. It can be observed that the best value is obtained with the fifth-order phase-integral approximation. It should be noticed that the   slow-roll approximation works well since  the parameters $\epsilon_1$, $\epsilon_2$ and $\delta_n$ are small.

\begin{table}[htbp]
\begin{center}
\begin{tabular}{c|c|c|c|c|c|c|c}
\hline
\hline

&num$^\text{\tiny{a}}$ &phi3$^\text{\tiny{b}}$ &  phi5$^\text{\tiny{c}}$ & sr1$^\text{\tiny{d}}$ & sr2$^\text{\tiny{e}}$  &   phi1$^\text{\tiny{f}}$, WKB$^\text{\tiny{g}}$, ua1$^\text{\tiny{h}}$ &  ua2$^\text{\tiny{i}}$\\
\hline
$P_\sca(k)\times 10^{-11}$ & $8.3300$ & $8.3419$ & $8.3287$ & $8.2282$ &                                           $8.2299$ & $7.4946$ & $8.2752$\\
$P_\ten(k)\times 10^{-12}$ & $1.6109$ & $1.6132$ & $1.6106$ & $1.6047$ &                                           $1.6049$ & $1.4486$ & $1.6004$\\
 $n_\sca(k)$               & $0.960473$ & $0.960472$ & $0.96473$ & $0.960700$ &                                  $0.960491$ & $0.960453$ & $0.960486$\\
 $n_\ten(k)$               & $-0.019965$ & $-0.019982$ & $-0.019965$ & $-0.019650$ &                              $-0.019948$ & $-0.020000$ & $-0.019984$\\
\hline\hline
\end{tabular}
\end{center}
\caption{\small{Value of $P_\sca(k)$, $P_\ten(k)$, $n_\sca(k)$, and $n_\ten(k)$ obtained  with different approximation methods for the chaotic inflationary model  $\frac{1}{2}m^2\phi^2$ for the mode  $k=0.05\,\Mpc^{-1}$.}}
\label{ch2_t1}
\begin{tabular*}{0.8\textwidth}{l}
$^\text{\tiny{a}}$\small{Numerical}\\
$^\text{\tiny{b}}$\small{Third-order phase-integral approximation}\\
$^\text{\tiny{c}}$\small{Fifth-order phase-integral approximation}\\
$^\text{\tiny{d}}$\small{First-order slow-roll approximation}\\
$^\text{\tiny{e}}$\small{Second-order slow-roll approximation}\\
$^\text{\tiny{f}}$\small{First-order phase integral approximation}\\
$^\text{\tiny{g}}$\small{WKB approximation}\\
$^\text{\tiny{h}}$\small{First-order uniform approximation}\\
$^\text{\tiny{i}}$\small{Second-order improved uniform approximation}\\
\end{tabular*}
\end{table}

\begin{figure}[htbp]
\begin{center}
\includegraphics[scale=0.465]{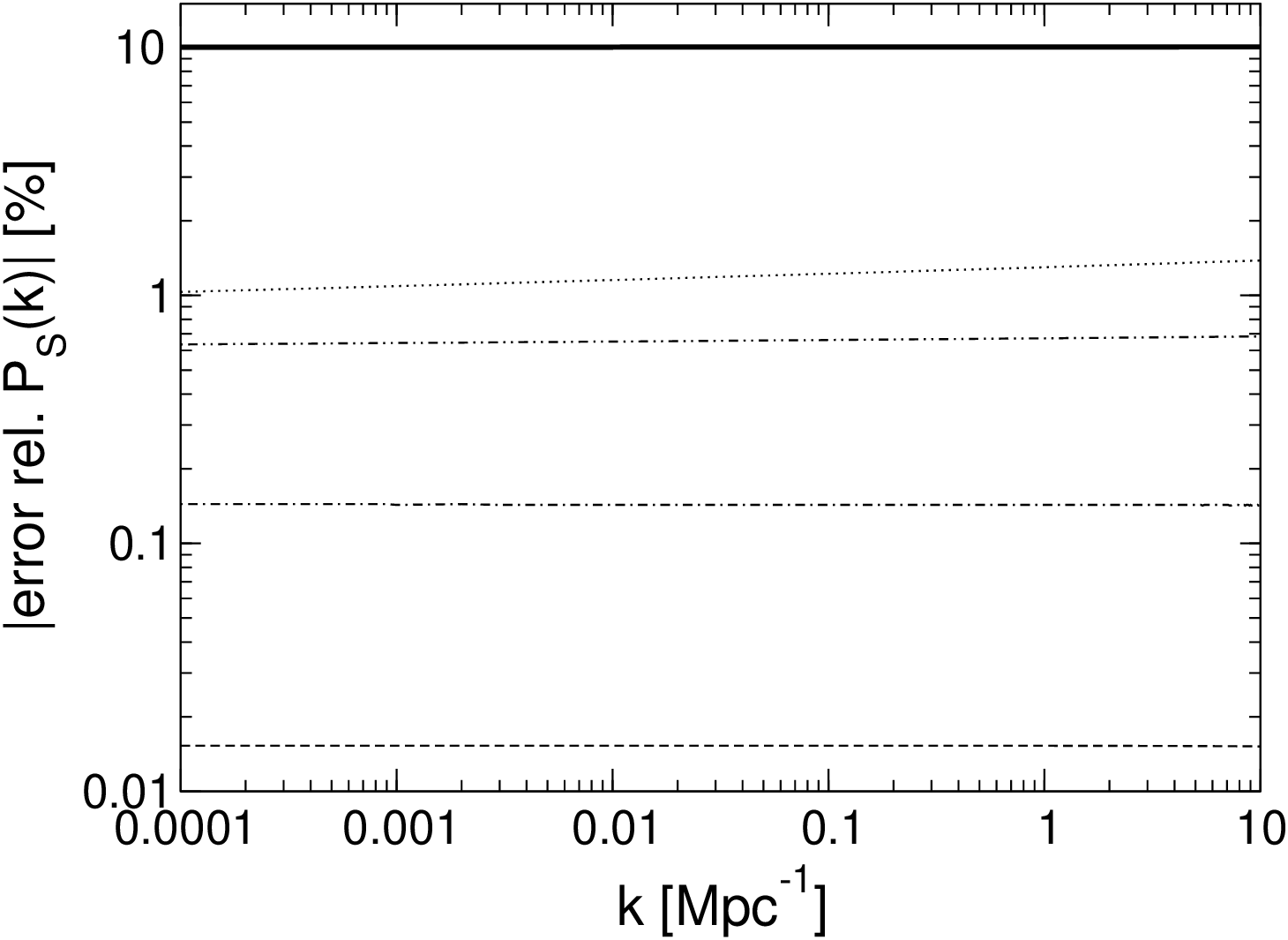}
\caption{\small{Relative error for $P_\sca(k)$  for the chaotic inflationary $\frac{1}{2}m^2\phi^2$ model. Dot-dashed line: third-order phase-integral approximation; dashed line: fifth-order phase-integral approximation; thick solid line: first-order phase-integral approximation; WKB and  first-order uniform approximation; dashed double-dots line: second-order improved uniform approximation; dotted line: first and second-order slow-roll approximation. The inset is an enlargement of the figure.}}
\label{ch2:error_PS}
\end{center}
\end{figure}

\begin{figure}[htbp]
\begin{center}
\includegraphics[scale=0.465]{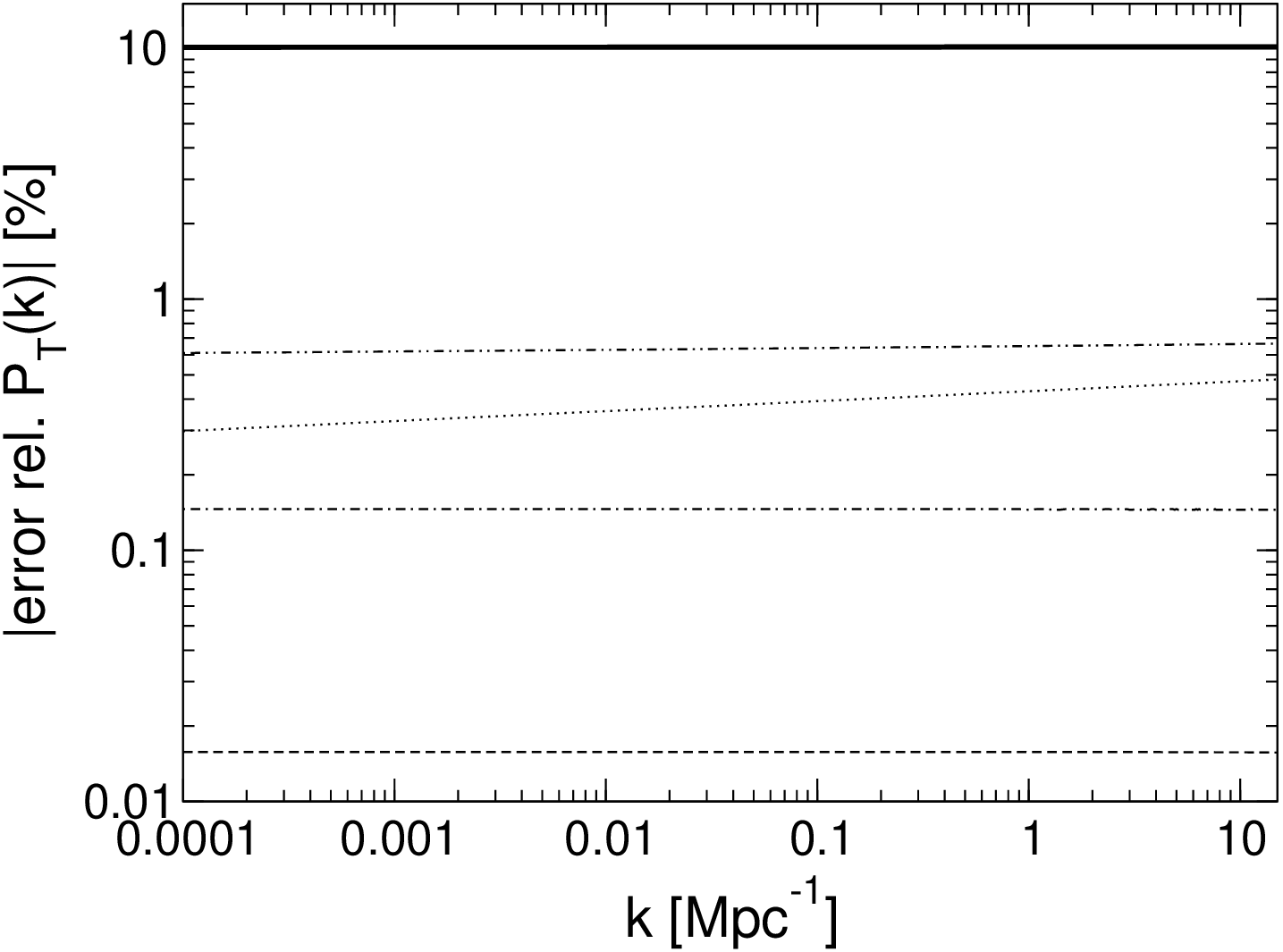}
\caption{\small{Relative error for $P_\ten(k)$  for the chaotic inflationary $\frac{1}{2}m^2\phi^2$ model. Dot-dashed line: third-order phase-integral approximation, dashed line: fifth-order phase-integral approximation; thick solid line: first-order phase-integral approximation; WKB and  first-order uniform approximation; dashed double-dots line: second-order improved uniform approximation; dotted line: first and second-order slow-roll approximation. The inset is an enlargement of the figure.}}
\label{ch2:error_PT}
\end{center}
\end{figure}

\begin{figure}[htbp]
\begin{center}
\includegraphics[scale=0.465]{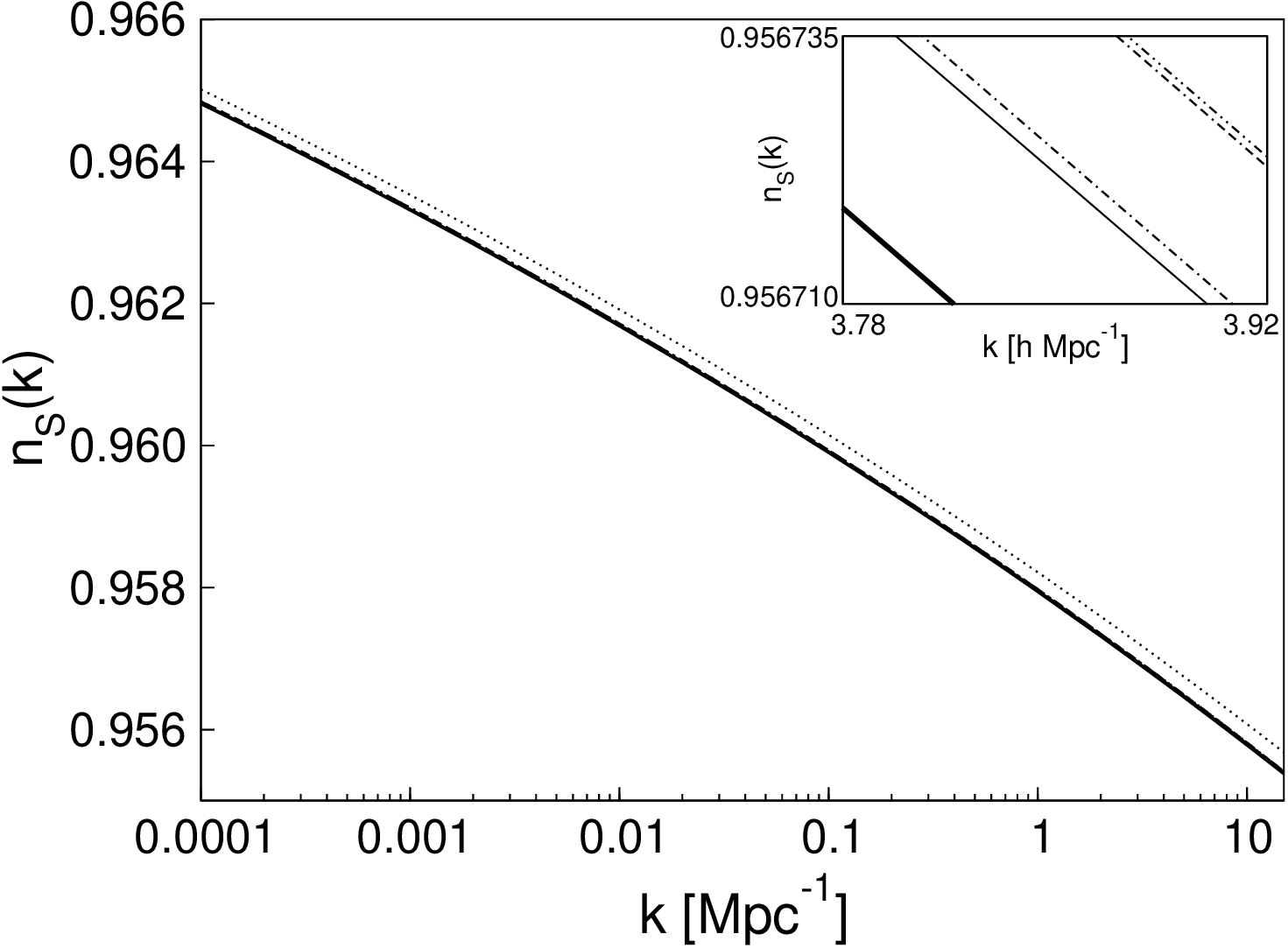}
\caption{$n_\sca(k)$  for the chaotic inflationary $\frac{1}{2}m^2\phi^2$ model.  Thin solid line: numerical result (ICs phi3) and fifth-order phase-integral approximation; dot-dashed line: third-order phase-integral approximation;thick solid line: first-order phase-integral approximation; WKB and  first-order uniform approximation; dashed double-dots line: second-order improved uniform approximation; double-dashed dot line: second-order slow-roll approximation; dotted line: first-order slow-roll approximation. The inset is an enlargement of the figure.}
\label{ch2:nS}
\end{center}
\end{figure}

\begin{figure}[htbp]
\begin{center}
\includegraphics[scale=0.465]{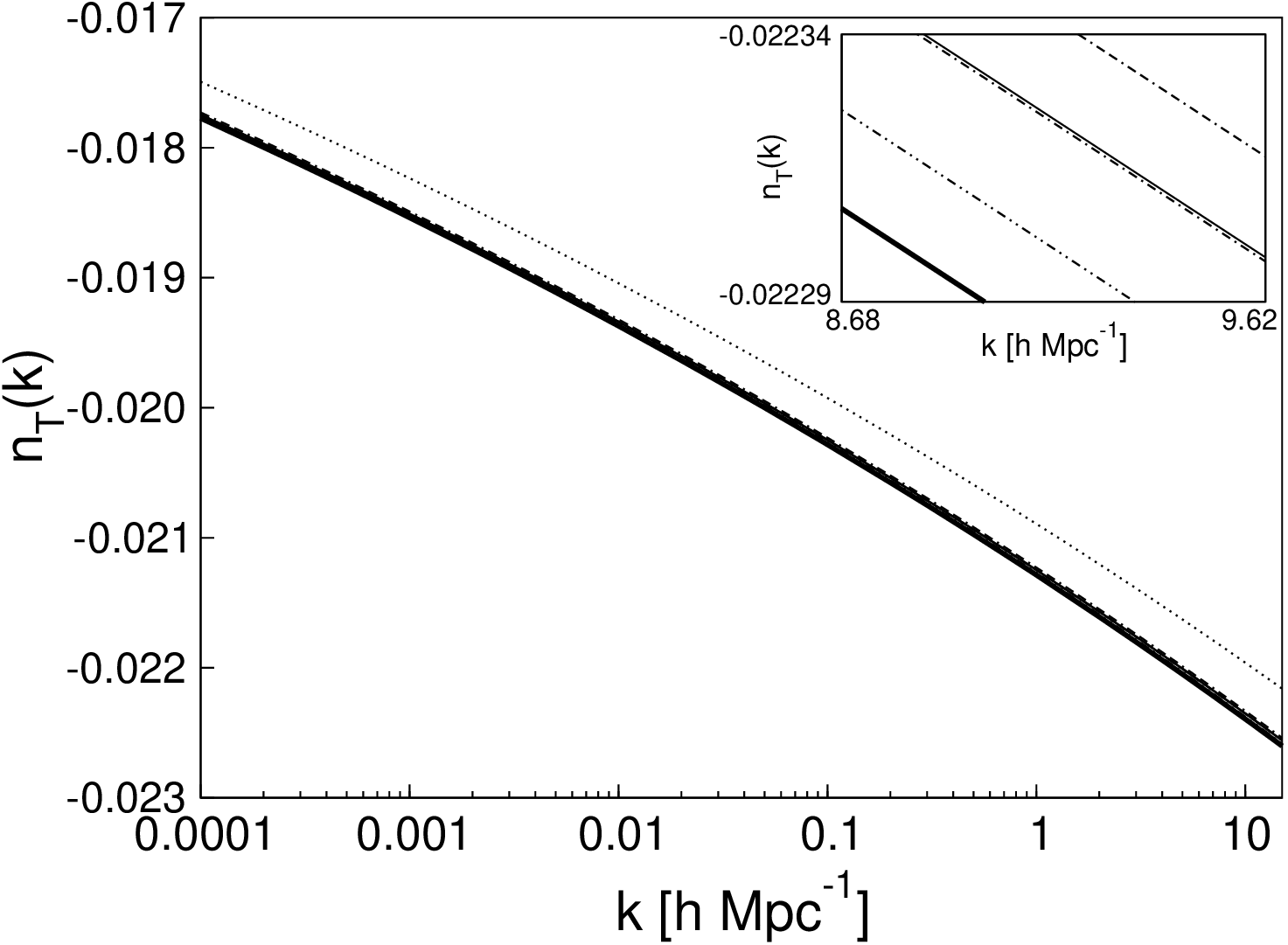}
\caption{$n_\ten(k)$  for the chaotic inflationary $\frac{1}{2}m^2\phi^2$ model.  Thin solid line: numerical result (ICs phi3) and fifth-order phase-integral approximation; dot-dashed line: third-order phase-integral approximation; thick solid line: first-order phase-integral approximation; WKB and  first-order uniform approximation; dashed double-dots line: second-order improved uniform approximation; double-dashed dot line: second-order slow-roll approximation; dotted line: first-order slow-roll approximation. The inset is an enlargement of the figure.}
\label{ch2:nT}
\end{center}
\end{figure}

Figures \ref{ch2:error_PS} and \ref{ch2:error_PT} show the relative error with respect to the numerical result that is obtained using the expression

\begin{equation}
\text{error rel.}\,\,P_{\sca,\ten}(k) = \frac{\left[P_{\sca,\ten}^\text{\text{approx}}(k)-P_{\sca,\ten}^\num(k)\right]}{P_{\sca,\ten}^\num(k)}\times 100.
\end{equation}
The first-order phase integral approximation, the WKB and the first-order uniform approximation give the same result, and deviate from the numerical result in  $10\%$. The second-order improved uniform approximation gives an error of $0.6\%$. With the first and second-order slow-roll approximation we have an error of $1\%$ for $P_\sca(k)$ and of $0.4\%$ for $P_\ten(k)$.  Using the third-order phase-integral approximation the error gives $0.15\%$, whereas the fifth-order phase-integral reduces to  $ 0.015\%$ in both cases. Fig. \ref{ch2:nS} and Fig.  \ref{ch2:nT} show the results for the  spectral indices $n_\sca(k)$ and  $n_\ten(k)$  respectively.



\section{Concluding remarks}

The results reported in this article show that, in comparison with
other approximation methods, the  phase integral approach gives very
good results  for the scalar and tensor spectra in the quadratic
inflationary model. The phase-integral approximation gives very
accurate results as soon as the the integral $\mu(z,z_{0})$ is
small. Figures \ref{ch2:Re(uk)}-\ref{ch2:Abs(vk)} show that the
phase integral approximation fails in the vicinity of the turning
point $-\nu$, range where the $\mu$-integral diverges. The selection
of the base function $Q(z)$ guarantees that $\mu\ll 1$ far from the
turning point at any order of approximation.  Since the scalar and
tensor power spectra as well as the spectral indices are evaluated
as $-k\eta\rightarrow 0$, the limit  is taken far from the horizon
(turning point), therefore their computation is not affected by the
presence of the turning point.

Since the WKB method can be regarded as a first-order approximation
of the  phase-integral approximation with $Q^2(z)=R(z)$, it should
be expected that the phase-integral method works in those cases
where the WKB methods gives good estimates and slow-roll fails, that
is the case where inflation is generated by a chaotic potential with
a step \cite{Hunt,Casadio2006}. The good agreement between the
numerical results and those obtained with the phase-integral
approximation shows that the phase integral method is a very useful
approximation tool for computing the scalar and tensor the power
spectra in  a wide range of inflationary scenarios.

\acknowledgments

One of the authors (CR) wishes to express her gratitude to  Carlos
Cunha  for enlightening discussions and for his help in the
implementation of the numerical code for solving the perturbation
equations. We thank Dr. Ernesto Medina for reading and improving the
manuscript. This work was partially supported by FONACIT under
project G-2001000712.


\end{document}